\documentclass{aa} %longauth
\usepackage{mathtools}
\usepackage{graphicx}
\usepackage{txfonts}
\usepackage{natbib}
\usepackage{longtable}
\usepackage{soul}

\def\msol{$M_{\odot}$}
\def\micron{$\mu$m}

\def\her{\emph{Herschel}}

\begin{document}

\title{How the power spectrum of dust continuum images may hide the presence of a characteristic filament width }

\author{A. Roy\inst{\ref{inst1},\ref{inst8}}
  \and Ph. Andr\'{e} \inst{\ref{inst1}}
  \and D. Arzoumanian\inst{\ref{inst2}}
%  \and N. Peretto \inst{\ref{inst-cardif}}
    \and M.-A. Miville-Desch\^{e}nes\inst{\ref{inst1},\ref{inst3}}
    \and V. K\"{o}nyves\inst{\ref{inst1}}
     \and N. Schneider\inst{\ref{inst4}} 
        \and S. Pezzuto\inst{\ref{inst5}} 
      \and P. Palmeirim\inst{\ref{inst6}}        
    \and  J. M. Kirk\inst{\ref{inst7}}
%  \and A. Men'shchikov \inst{\ref{inst1}}    
%\and
%	the Herschel Gould Belt Survey Team
    }

\institute{Laboratoire d'Astrophysique (AIM), CEA, CNRS, Universit\'{e} Paris-Saclay, Universit\'{e} Paris Diderot, Sorbonne Paris Cit\'{e}, 91191 Gif-sur-Yvette, France\label{inst1} 
\and  Laboratoire d'Astrophysique de Bordeaux, Univ. Bordeaux, CNRS, B18N, all\'ee G. Saint-Hilaire, 33615 Pessac, France\label{inst8}
\and Department of Physics, Nagoya University, Furo-cho, Chikusa-ku, Nagoya, Aichi, 464-8602, Japan\label{inst2}
\and Institut d'Astrophysique Spatiale, CNRS, Univ. Paris-Sud, Universit\'e Paris-Saclay, B\^atiment 121, 91405 Orsay cedex, France\label{inst3} 
\and I. Physik. Institut, University of Cologne, Z\"ulpicher Str. 77, 50937 Koeln, Germany  \label{inst4}
\and INAF - Istituto di Astrofisica e Planetologia Spaziali, via Fosso del Cavaliere 100, 00133, Roma, Italy  \label{inst5}
\and Instituto de Astrof\'isica e Ci{\^e}ncias do Espa\c{c}o, Universidade  
do Porto, CAUP, Rua das Estrelas, PT4150-762 Porto, Portugal \label{inst6} 
\and University of Central Lancashire, Preston, Lancashire, PR1 2HE,
United Kingdom \label{inst7}
\and E-mails: aroy@cita.utoronto.ca, pandre@cea.fr}
%Arabindo.Roy@u-bordeaux.fr, pandre@cea.fr}

\titlerunning{Effect of a characteristic filament width on the power spectrum of cloud images}

\abstract
{{\it Herschel} observations of interstellar clouds support a paradigm for star formation in which molecular filaments play a central role.
One of the foundations of this paradigm is the finding, based on detailed studies of the transverse column density profiles 
observed with {\it Herschel}, that nearby molecular filaments share a common inner width of $\sim 0.1\,$pc. 
The existence of a characteristic filament width has been recently questioned, however, on the grounds that it seems inconsistent 
with the scale-free nature of the power spectrum of interstellar cloud images.
}
{In an effort to clarify the origin of this apparent discrepancy, we examined the power spectra of the {\it Herschel}/SPIRE $250\, \mu$m 
images of the Polaris, Aquila, and Taurus--L1495 clouds in detail and performed a number of simple numerical experiments by injecting synthetic filaments 
in both the {\it Herschel} images and synthetic background images. 
}
{We constructed several populations of synthetic filaments of $0.1\,$pc width with realistic area filling factors ($A_{\rm fil}$)
and distributions of column density contrasts ($\delta_c$). After adding synthetic filaments to the original {\it Herschel} images, 
we re-computed the image power spectra and compared the results with the original, essentially scale-free power spectra. 
We used the $\chi^2_{\rm variance}$ of the residuals between the best power-law fit  and the output power spectrum in each simulation 
as a diagnostic of the presence (or absence) of a significant departure from a scale-free power spectrum.
}
{We found that $\chi^2_{\rm variance}$ depends primarily on the combined parameter $\delta_c^2\, A_{\rm fil}$. 
According to our numerical experiments, a significant departure from a scale-free behavior and thus the presence of a 
characteristic filament width become detectable in the power spectrum when $\delta_c^2\, A_{\rm fil} \gtrapprox  0.1$ for synthetic filaments with Gaussian profiles 
and  $\delta_c^2\, A_{\rm fil} \gtrapprox  0.4$ for synthetic filaments with Plummer-like density profiles.
Analysis of the real {\it Herschel} $250\, \mu$m data suggests that  $\delta_c^2\,A_{\rm fil}$ is $\sim 0.01$ in the case of the Polaris cloud 
and $\sim 0.016$ in the Aquila cloud, significantly below the fiducial detection limit of $\delta_c^2\, A_{\rm fil} \sim 0.1$ in both cases. 
In both clouds, the observed filament contrasts and area filling factors are such that the filamentary structure contributes only $\sim 1/5$ 
of the power in the image power spectrum at angular frequencies 
%$k \la k_{\rm fil}$ 
where an effect of the characteristic filament width is expected. 
}
{We conclude that the essentially scale-free power spectra of {\it Herschel} images remain consistent with the existence 
of a characteristic filament width $\sim 0.1\,$pc and do not invalidate the conclusions drawn from studies of the filament profiles.
}
%}

\maketitle

\section{Introduction}
Recent {\it Herschel} imaging observations of nearby molecular clouds, 
%especially 
e.g., those obtained as part of the {\it Herschel}
Gould Belt Survey (HGBS; \citealp{andre2010}), indicate that filamentary structures
%invariably maintain a quasi universal 
%tend to 
are characterized by a common inner width $W_{\rm fil} \sim$ 0.1 pc, with only a factor of $\sim 2$ spread around 
this value, over a wide range of column densities \citep[][]{arzoumanian2011, arzoumanian2017, rosolowsky2015}.  
If confirmed, the existence of such a characteristic filament width has remarkable implications for the star formation process 
and is one of the bases of a proposed filamentary paradigm for solar-type star formation \citep{andre2014}.
In particular, it may set a critical column density threshold above which most stars form in filamentary molecular clouds.
%\citep{andre2014,shimajiri2017}.  
For filaments of $\sim 0.1\,$pc width and a typical gas temperature of 10 K, the critical mass per unit length 
$M_{\rm line, crit} = 2\, c_s^2/G \sim 16\, $ \msol/pc \citep[cf.][]{Inutsuka1997}
%(cf. Inutsuka \& Miyama 1997) 
indeed translates to a critical column density  $\Sigma_{\rm gas,crit} \sim M_{\rm line, crit} /W_{\rm fil} \sim 160 $\msol/pc$^{2}$, 
which is close to the background column density threshold above which prestellar cores are found with {\it Herschel}  in nearby 
regions \citep[e.g.][]{konyves2015,marsh2016}. Above this threshold, the star formation rate is observed to be directly proportional 
to the mass of dense molecular gas in both nearby clouds and external galaxies \citep[e.g.][]{Gao2004,Heiderman2010,lada2010,shimajiri2017}.
%empirical finding of 130 \msol/pc$^{2}$ by \cite{Heiderman2010}, for external galaxies.

%The primary basis for 
\citet{arzoumanian2011} suggested the existence of a characteristic filament width 
after fitting simple Gaussian or Plummer-like model profiles to the transverse profiles 
observed with {\it Herschel} for a broad sample of nearby filaments (see also \citealp{arzoumanian2017}).
In the analysis of Arzoumanian et al., thermally supercritical filaments (with $M_{\rm line} > 2c_s^2/G$) tend to have Plummer-like density
profiles with a flat inner region of radius $R_{\rm flat} $ and a decreasing power-law wing $\rho \propto r^{-2}$ at larger radii.
%with an exponent similar to the inverse square.  
In contrast, low column density, thermally subcritical filaments (with $M_{\rm line} < 2c_s^2/G$) tend to be better 
described by Gaussian density profiles. 
%do not have power-law profiles.  They arerather best described by Gaussian profiles instead.

Why molecular filaments seem to share such a characteristic width is still a debated theoretical problem 
\citep[e.g.][]{hennebelle2013, fischera2012, federrath2016, auddy2016}. 
In order to ascertain whether the presence of this possibly universal filament scale is robust, 
the observational data also need to be investigated using various other means.  
%Toward this aim,
In a recent paper, \cite{panopoulou2017} tested the possibility of identifying a
characteristic scale using a power spectrum analysis.  In their study
they argued that, had there been a characteristic filament width, its signature
should have manifested itself in the power spectrum of {\it Herschel} images of nearby clouds, 
%interstellar medium (ISM) images,  
either as a kink or as a change in slope at an angular frequency
corresponding to the characteristic scale.

In the present paper, we revisit the latter issue from an observer's standpoint,
exploring the parameter space with realistic filament properties
consistent with the observational data, in particular taking into account realistic distributions  
of filament contrasts and area filling factors. 
%We have also investigated
%the role of the combined effect of density contrasts and area filling factor  
%and their products
%$\delta 
%choice of
%filament profiles according to local background column
%densities. 
%Previous observations have shown that for low column
%density filaments in a non-star forming cloud ( e.g., filaments in the
%Polaris flare translucent cloud) the average peak filament contrast,
%$\delta_c = (I^{\rm peak} - I^{\rm bkg})/I^{\rm bkg}$, with respect to
%the background was around $\sim$ 1 but with extremely low area  (Arzoumanian et al. 2017, {\it in
%  prep}).
%
%In the present analysis we have selected 
To this end, we selected two extreme regions imaged by the HGBS, 
namely the Polaris translucent cloud, mainly dominated by low density subcritical ($M_{\rm line} < M_{\rm line,crit}$)
filaments, and the Aquila complex, which contains a fair population of high
column density supercritical ($M_{\rm line} > M_{\rm line,crit}$) filaments. 
We also used the B211/B213 field in the Taurus cloud, which is dominated by a single, 
marginally supercritical filament.

The layout of the paper is as follows. In Sect.~\ref{sec:construct-fil} we describe the construction of synthetic filament images and their power spectra. 
%In Sect. \ref{sec:diagnostic} 
We also develop a diagnostic for the detection of a characteristic filament width in a power spectrum plot.  
In Sect.~\ref{Sec:polaris} and Sect.~\ref{Sec:Aquila}, 
we perform a power spectrum analysis of the {\it Herschel} images of the 
Polaris and Aquila clouds, respectively,
and compare the results to those obtained on synthetic maps 
after adding simulated filaments. 
 In Sect.~\ref{sect:taurus},  we compared power spectra of a sub-region of Taurus molecular cloud encompassing the Taurus main filament to a  synthetic filament with similar physical properties as B211/B213. 
In Sect. \ref{sec:deltaAfil}, 
we investigate the combined effect of filament column density contrast ($\delta_c$)  
and area filling factor ($A_{\rm fil}$). 
Finally, we summarize our results in Sect. \ref{Sec:conclusion}.

\begin{figure}
%  \centering
  \resizebox{0.9\hsize}{!}{\includegraphics[angle=0]{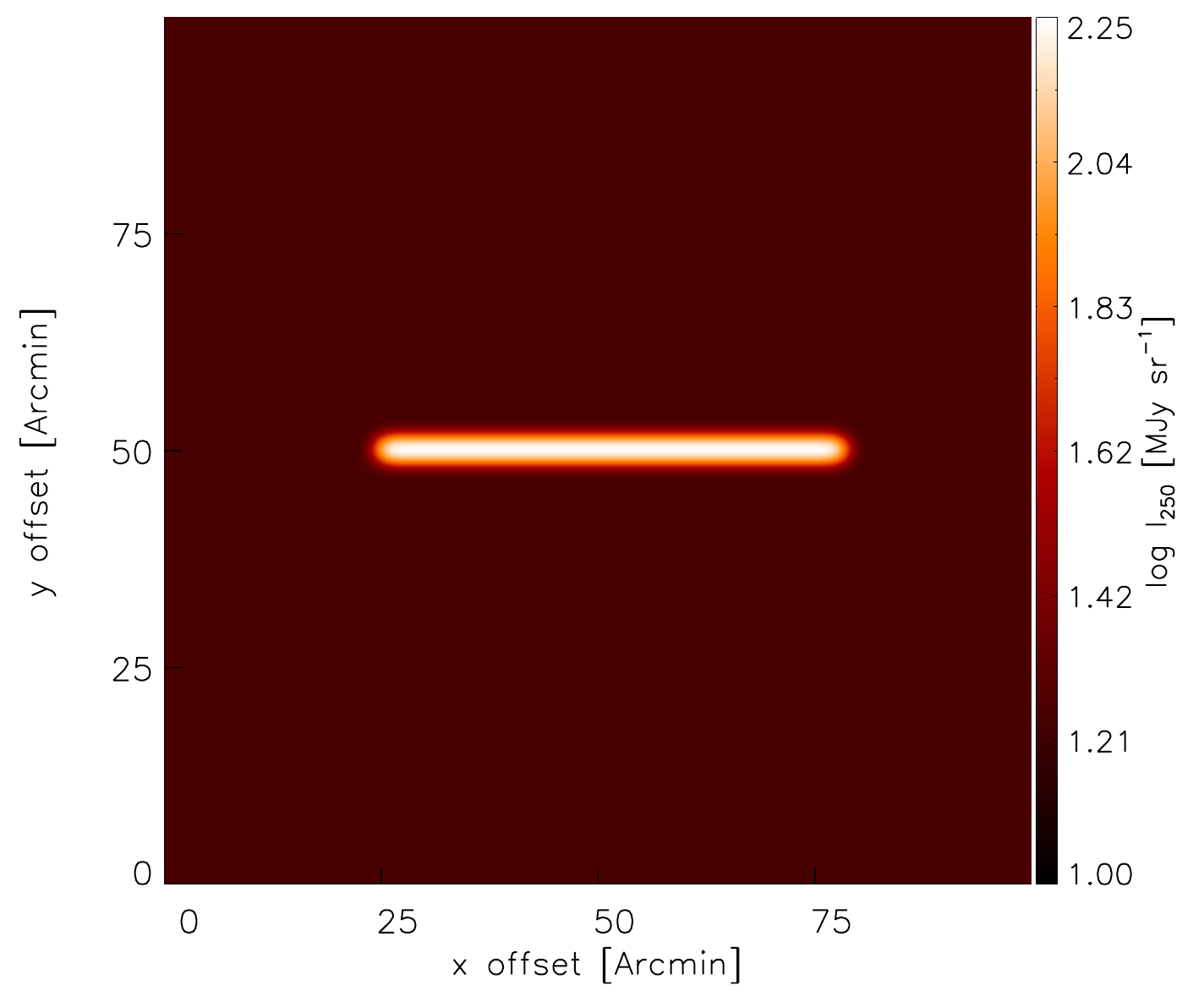}}
 \caption{ Image of a simulated filament with a Gaussian transverse profile and a
%   flat inner 
   FWHM width of 0.1 pc, projected at a distance of 140 pc.  Here, the
   level of filament contrast was adjusted so that $\delta_{\rm c} \sim$  10.  }
\label{fig-gauss-fil}
\end{figure}

\section{Construction of synthetic filaments and their power spectra}\label{sec:single-filament} \label{sec:construct-fil}
Figure~\ref{fig-gauss-fil} shows an example of a synthetic filament with
a transverse Gaussian profile and a projected spatial inner width (FWHM) of
0.1 pc at a distance of 140 pc.  Mathematically, the 2D image of a filament with a
Gaussian profile can be expressed as
\begin{equation}
I_{\rm Cylinder}(x, y) = C\,[\delta^L(ax+by+c)\times \Pi_{L}] \star G_{\theta_{\rm fil}}(x,y),    
\label{eq:cylinder}
\end{equation}
where, $C$ is the amplitude factor (related to the filament contrast) of the 
delta line function $\delta^L(ax+by+c)$, $\Pi_L$ is a rectangle function, 
and the $\star$ symbol denotes the convolution operator. 
The delta line function assumes a value of unity when its parameter $ax+by+c = 0$, and zero
elsewhere. The expression $ax+by+c = 0$ is the equation of a straight
line where the $a$, $b$ coefficients determine the slope of the straight line and $c$
is the intercept.
The rectangle function, $\Pi_{L}$, which has a value of unity over the length $L$ of the line function and zero elsewhere, transforms the line into a line-segment of length $L$.  
In order to make a Gaussian filament profile, we convolve the entire line segment [$\delta(ax+by+c)\times \Pi_L$] with a Gaussian kernel, $G_{\theta_{\rm fil}}(x,y)$, of 
full width at half maximum $\rm{FWHM} =\theta_{\rm fil}$. 
%The symbol $\star$ denotes the convolution operator.

The FWHM of the Gaussian kernel $=\theta_{\rm fil}$ is chosen such
that the projected spatial width is $W_{\rm fil}$.  To create a
characteristic $W_{\rm fil}$ inner width of a filament at a distance $d$,  the required $\theta_{\rm fil}$ is
\begin{equation}
\theta_{\rm fil} \simeq 147\arcsec \times \left(\frac{W_{\rm fil}}{0.1~ \rm pc} \right) \times \left(\frac{140~ \rm pc}{d} \right).
\end{equation}
The choice of the parameter $C$ depends upon the required level of
filament contrast defined as $\delta_{\rm c} = (I^{\rm peak} - I^{\rm bkg})/I^{\rm bkg} $ and on the dilution factor of the
convolution kernel,
\begin{equation}
C \approx \delta_{\rm c} \theta_{\rm fil}.
\label{eq:C}
\end{equation}   
For example, in order to create a Gaussian filament profile of spatial
FWHM $=$ 0.1 pc and contrast $\delta_c = 0.4$ at the distance of
Polaris ($d = 140$ pc, \citealp{falgarone1998}), we used a Gaussian convolution kernel of
$\theta_{\rm fil} \sim$ 147\arcsec (see Eq.\ref{eq:C}), and a contrast
amplitude of $C\sim 0.4 \times 147\arcsec/\theta_{\rm pix}$, where $\theta_{\rm pix}$ is the pixel size of the image. 
We adopted a pixel size
%Nyquist sampled  
$\theta_{\rm pix}$=6\arcsec\  for {\it Herschel}/SPIRE 250~\micron\ images (18.2\arcsec\ beam  resolution).

The Fourier transform of a 
%the 
2D image 
%of a cylindrical image 
can be expressed as
\begin{equation}
\hat{I}(k_x,k_y) = \int I(x,y)e^{-2\pi i (k_xx + k_yy)} dxdy,  
\label{eq:fourier}
\end{equation}
where $dx\,dy$ is the infinitesimal surface area, and $\int dx\,dy=S$ is the total surface area, $S$, covered by the map.
For an image of a single cylindrical filament,  most of the contribution to the integral in Eq.~\ref{eq:fourier} 
comes from integration over the central part of the filament which encloses 75\% of the total intensity fluctuations.

The power spectrum of a cylindrical intensity distribution can be
written analytically as
\begin{eqnarray}
P_{cylinder}(k_x, k_y) &= &|\hat{I}(k_x,k_y)|^2, \nonumber\\
                        &=&| FT(\delta^L(ax+by+c)(k_x,k_y) |^2 \, \hat{G}^2(k_x,k_y), \nonumber\\
                       &=& |\hat{\delta}(bk_x -ak_y)|^2 \,  \hat{G}^2(k_x,k_y),
\label{eq:ft}
\end{eqnarray}     
where $\hat{\delta}(bk_x -ak_y) $ is the Fourier transform of the delta  line function $\delta^L(ax+by+c)$ 
and $\hat{G}(k_x,k_y)$ is the Fourier transform of the
convolution kernel. The power spectrum of the Gaussian kernel, $\hat{G}^2(k_x,k_y)$,  is also a
 Gaussian, and its FWHM width $\Gamma_{\rm fil}$ is related to the FWHM width
  $\theta_{\rm fil}$ of $G_{\theta_{\rm fil} }(x,y)$ through the relation:  
  \begin{eqnarray}
  \Gamma_{\rm fil} &=& \sqrt{8}{\rm ln2}/\pi\theta_{\rm fil} \nonumber\\
  &\sim& 0.6/\theta_{\rm fil}.
  \end{eqnarray}
%.
%In theory, 
One may thus expect a characteristic filament width $\theta_{\rm fil}$ to lead 
to a signature in the power spectrum at angular frequencies
%correspond to an angular frequency $k_{\rm fil}$ = $\Gamma_{\rm fil}$. 
$k_{\rm fil} \sim \Gamma_{\rm fil}$. 

\begin{figure}
  \resizebox{ 0.95\hsize}{!}{\includegraphics[angle=0]{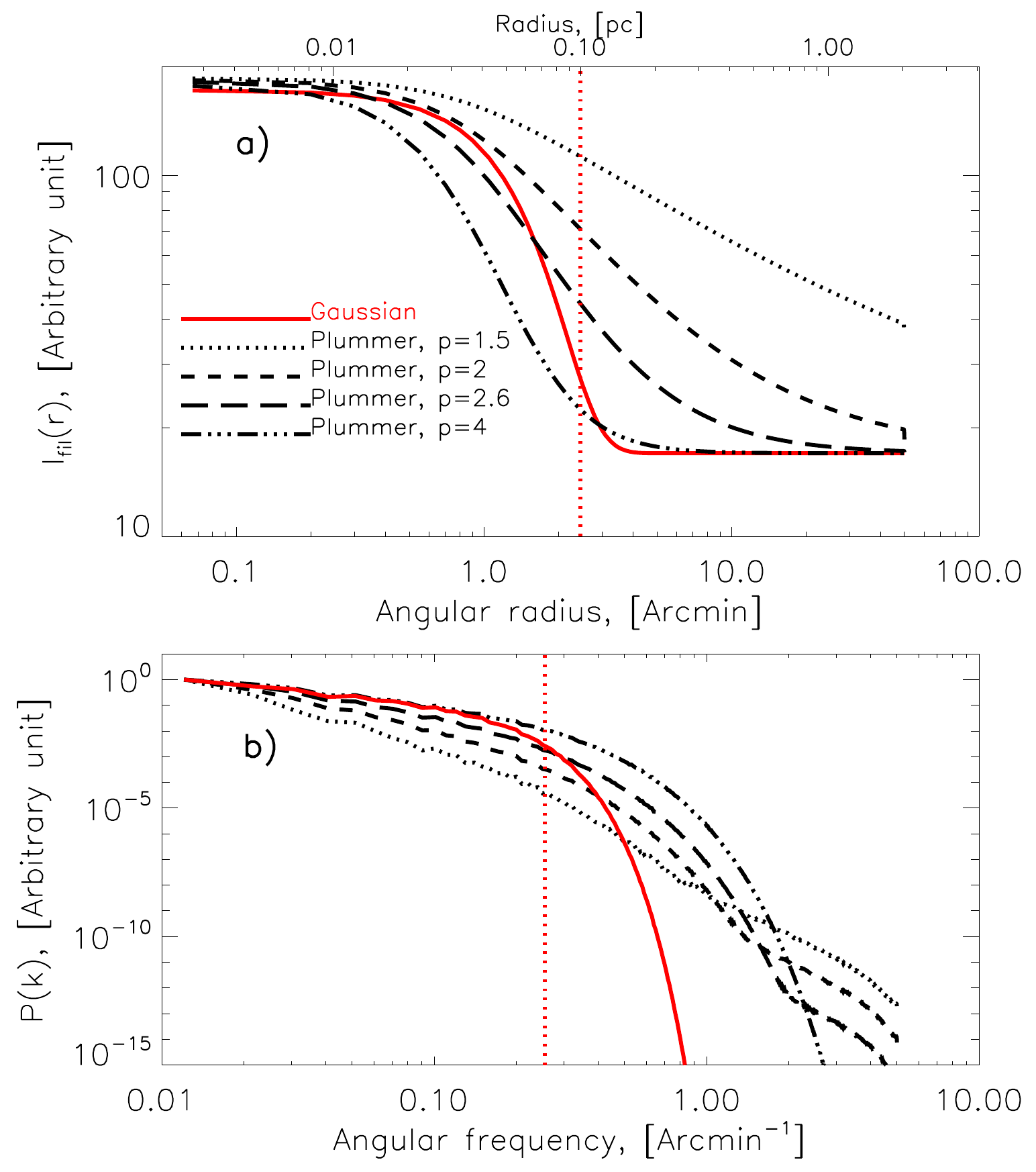}}%{Pipe_fig.ps}}
 \caption{Transverse profiles of several simple model filaments {\bf (a)} and corresponding power spectra {\bf (b)}.
In panel {\bf a)}, the red curve  shows the Gaussian column density profile of the filament displayed in Fig.~\ref{fig-gauss-fil},
which has a FWHM width of 0.1 pc at a distance of 140 pc. The red vertical line marks the FWHM of the Gaussian profile.
The black dashed and dotted curves display 
Plummer-like filament profiles with $R_{\rm flat}\sim 0.03\,$pc and logarithmic slopes  p=1.5, 2, 2.6, and 4.    
%Fil constant $R_{\rm flat}\sim 0.03 pc$      
%
In {\bf b)}, all power spectra were normalized to 1 at the lowest angular frequency $k_{\rm min} = \frac{1}{{\rm Map\, Size}}$. 
%Power spectra of synthetic filament images whose profiles are shown in the upper panel.  
%For all of the profiles,
The power spectra all decrease sharply at  high angular frequencies, with 
the highest rate of decrease obtained for the Gaussian model (red curve).  
For the Plummer models, the rate of decrease is higher for higher $p$ values. 
Note the kink near $k$= 0.8 arcmin$^{-1}$ for the Plummer model with $p = 1.5$, which disappears for higher $p$ values. 
The red vertical dashed line denotes scale $\Gamma_{\rm fil}$= 0.24 arcmin$^{-1}$.  }
   \label{fig-kernel-ps}
\end{figure}

For multiple, randomly distributed filaments, as in the simulations discussed below, it is not possible 
to obtain the power spectrum analytically.  
We therefore used the IDL-based routine FFT to compute the power spectrum.
Nevertheless, Eq.~\ref{eq:ft} is useful to appreciate
how the power spectrum of an image with a single filament is dominated by the power
spectrum of the convolution kernel. 
As an illustration, Fig.~\ref{fig-kernel-ps}b displays the power spectra of images including a single model filament 
with either a Gaussian or a Plummer-like density profile.
%panels a) and b) of Fig.~\ref{fig-kernel-ps} display model Gaussian and Plummer-like filament profiles and their corresponding power spectra.   
The red curve in Fig.~\ref{fig-kernel-ps}a shows the radial profile of the Gaussian
filament displayed in Fig.~\ref{fig-gauss-fil}. The over-plotted black curves show the profiles of filaments 
featuring Plummer-like power-law wings at large radii, with power-law slopes ranging from $p = 1.5$ to $p = 4$. 
%of p=1.5, 2, 2.6 and 4.
The flat inner region of each Plummer-like model filament had 
%power-law wing profile has
a constant $R_{\rm flat }$ of $\sim$ 0.03 pc.
An example of filament with a transverse Plummer profile is shown in Fig~\ref{fig-plumm-fil}.  
%For p=2 profile, the inner region of Plummer profile can be approximately fitted with a Gaussian function
%of FWHM = 3$R_{\flat}$.
Figure ~\ref{fig-kernel-ps}b displays the power spectra corresponding to the filament profiles shown in Fig.~\ref{fig-kernel-ps}a. 
At high angular frequencies, the power decreases exponentially following the same trend as the power spectrum of the convolution kernel. 
%Similarly, the black curves in show
%the power spectra of Plummer-like filaments with various power-law wings, with slopes of $p$= 1.5, 2, 2.6, and 4.  
An example of a filament with a Plummer-like radial profile is shown in
Fig~\ref{fig-plumm-fil}.

%\subsection{ Diagnostics for determining a characteristic width from the image power spectrum} \label{sec:diagnostic}
\subsection{Diagnosing the presence of a characteristic filament width from image power spectra} \label{sec:diagnostic}

%Far-infrared and submillimeter dust continuum images of the diffuse, cold interstellar medium (ISM) can be 
%decomposed into two components, one associated with emission from the diffuse background and the other 
%with a population of embedded filaments:
%modeled as a sum of emission from a diffuse background and a population of embedded filaments distribution:
In general, dust continuum images of the diffuse, cold interstellar medium (ISM) are well described by power-law power spectra, often attributed to the turbulent nature of the flow. {\it Herschel} images are also revealing a wealth of filaments. In the following we assume that these two contributions to the emission can be treated separately, in real space
\begin{equation} 
I_{\rm ISM }(x,y) = I_{\rm bkg}(x,y) + I_{\rm fil}(x,y).
\end {equation} 
Under the assumption that the filaments are randomly oriented and are not correlated with the diffuse background, we can
express the total power spectrum as: 
\begin{equation}
P_{\rm ISM}(k) = P_{\rm bkg} (k) + P_{\rm fil}(k),
\label{eq:ismpk}
\end{equation}
where $P_{\rm ISM}(k) $ is the total power spectrum of the ISM, and $P_{\rm bkg}(k)$ and  $P_{\rm fil}(k)$ represent the power spectra 
of the diffuse background and filament population, respectively. 
It is fair to assume that the power spectrum of dust images of the diffuse ISM 
%images 
follows a power-law, 
$P_{\rm bkg}(k)\propto k^{\gamma}$  with $\gamma \sim -2.7$ \citep{mamd2010,falgarone1998,stutzki1998,Robitaille2014}. 
%\citep{mamd2010,falgarone1998,stutzki1998,green1993,Robitaille2014}. 
From Fig.~\ref{fig-kernel-ps} and Eqs.~5--6, it is clear that the contribution of a population of filaments with constant width $\theta_{\rm fil}$ 
to the total power spectrum is not confined to a narrow range of spatial frequencies, but rather follows a shallow power law at angular frequencies
lower than $\Gamma_{\rm fil}$.   
%Since purely based on observations  we do not have prior knowledge of $P_{\rm bkg} (k)$, deciphering the presence of a characteristic scale 
%from a power spectrum analysis is a difficult task. 

In order to better visualize the filament contribution, we fit a power-law to the total power spectrum, $P_{\rm ISM}(k)$,
and then inspect the residuals,
\begin{equation}
Res(k) = \left[P_{\rm best~fit}(k) - P_{\rm ISM}(k) \right]/ P_{\rm ISM}(k), 
\label{eq:res}
\end{equation}
as a function of angular frequency. 
        To quantify the magnitude of the deviation from the best power-law
        fit, we use the $\chi^2_{\rm variance}$ as our metric.  We
        calculate the variance of the residuals in the vicinity
        of $k_{\rm fil}$ where the contribution of filament power is expected to be maximum\footnote{Note that $P_{\rm fil}(k) $ in Fig. \ref{fig-kernel-ps} is almost flat at $k<k_{\rm fil}$ and drops rapidly at $k>k_{\rm fil}$. }. 
        We define the
        variance as
        \begin{equation}
        \chi^2_{\rm variance} = \Sigma_{k_{\rm min}} ^{1.5 k_{\rm
            fil}}Res(k)^2/N_{\rm freq}, 
             \label{eq:chi-sq}
        \end{equation}
        where 
        $Res(k)$ is the
        residual at angular frequency $k$ defined by Eq.~\ref{eq:res}, and 
        $N_{\rm freq}$ is the total number
        of frequency modes\footnote{$k_{\rm min} = \frac{1}{{\rm Map\, Size}}$ is the minimum angular frequency considered in the power spectrum.} between
        $k_{\rm min}$ and $1.5\times k_{\rm fil} $.     
          The upper bound in the above summation is set to $1.5\times k_{\rm fil}$ in a conservative sense, 
          since the power spectrum of constant-width filaments drops at $k$ $>$ $k_{\rm fil}$ (see Fig.~\ref{fig-kernel-ps}b), 
          and  $Res(k)$ is therefore dominated by the diffuse ISM contribution for $k$ $>$ $k_{\rm fil}$. 
In principle, the image power spectrum of a scale-free ISM will have residuals close to zero, 
%a power-law ISM power spectrum has residuals close to zero. 
and any significant deviation of the residuals from zero at $k<k_{\rm fil}$ will be primarily due to the power spectrum of filaments 
(see Eq.~\ref{eq:ismpk}). 
Thus, one expects $\chi^2_{\rm variance} \propto \sum \left[P_{\rm best~fit}(k) - P_{\rm ISM}(k)\right]^2 \propto \sum P_{\rm fil}(k)^2$.
Simple dimensional analysis of the Parseval relation between $P_{\rm fil}(k)^2$ and  $|I_{\rm fil}(x,y)|^2$
provides deeper insight into the connection between $\chi^2_{\rm variance}$ and
%physical 
observable parameters of the filament population: 
\begin{eqnarray}
\left[\chi^2_{\rm variance}\right] &=& \left[ \sum P_{\rm fil}(k)^2 \right] = \left[\int |I_{\rm fil}(x,y)|^2dxdy\right]^2, %\nonumber\\
 %                       &=&\phi(\delta_c^2A_{\rm fil}),  %\nonumber\\
                  %     &=& |\hat{\delta}(bk_x -ak_y)|^2  \hat{G}^2(k_x,k_y),
\label{eq:parseval}
\end{eqnarray}   
%
%begin{eqnarray}
%\chi^2_{\rm variance} $\propto$ x \nonumber\\ %\int |f(x,y)|^2dxdy, \nonumber\\
%                     $\propto$ x %\delta_c^2A_{\rm fil},
%\end{eqnarray}
where  $[x]$ in square brackets denotes the dimension of quantity $x$. 
Dimensional analysis thus suggests that $\chi^2_{\rm variance}$ must be a function
of $\delta_c^2A_{\rm fil}$:
%the following functional dependence for $\chi^2_{\rm variance}$: 
\begin{equation}
\chi^2_{\rm variance} = \phi(\delta_c^2A_{\rm fil}).
\label{eq:phi}
\end{equation}
The  
%factor of filament contrast 
$\delta_c^2$ dependence comes from exploiting Eqs. 1 and 3, while 
the area filling factor $A_{\rm fil} \equiv S_{\rm fil}/S$ dependence comes from the fact that only the effective area $S_{\rm fil}$ 
over which filaments are distributed contribute to the integral on the right-hand side of Eq.~\ref{eq:parseval}.
        For low $A_{\rm fil}$ and $\delta_{c}$ the variance
        is very small, while for high $A_{\rm fil}$ and/or high
        $\delta_{c}$ the variance metric  can be very high. 
        We will explore the $\chi^2_{\rm variance} - \delta_c^2A_{\rm fil}$ parameter space in more detail in Sect.~\ref{sec:deltaAfil} below.

The magnitude/amplitude of excess power in the ISM power spectrum $P_{\rm ISM}(k)$ 
relative to the best-fit power law model power spectrum at a characteristic 
        frequency $k_{\rm fil}$ depends upon the combined effect of
        the mean filament contrast in the image and the fractional area
        covered by the filaments, $A_{\rm fil}$.
        The area filling factor $A_{\rm fil}$ can be expressed as $A_{\rm fil} =       \Sigma_{i=1}^{N_{\rm fil}} L_{\rm i} \times W_{\rm fil}$/$S$,
        where $L_{i}$ is the length of the $i^{\rm th}$ filament, 
        $W_{\rm fil}$ the transverse filament width ($\sim$ 0.1 pc), $S$  the total area coverage of the image
        being analyzed, and $N_{\rm fil} $  the total number of 
        %synthetic 
        filaments 
        %injected 
        in the image. 

%Equations \ref{eq:fourier} and \ref{eq:ismpk}
%provide an insightful information about the dependence of $\chi^2_{\rm variance} $ metric on density contrast
%$delta_c$ and area filling factor $ A_{\rm fil}$:
% \begin{eqnarray}
%\hat{I}_{ISM}(k_x, k_y) &= & \int_{S}I_{\rm bkg }({\bf r} )e^{-2\pi i ({\bf k.r}) } {\bf dr}  + \int_{S^\prime} I_{\rm fil}({\bf r}^ \prime)e^{-2\pi i (\bf k.r)}{\bf dr^\prime}, \nonumber\\
%                        &\approx& \hat{I}_{\rm bkg}(k_x,k_y) + A_{\rm fil}\int_{S} I_{\rm fil}({\bf r})e^{-2\pi i ({\bf k.r})}{\bf dr^\prime}, 
%\label{eq:ismft}
%\end{eqnarray}     
%where,  $\int_{S\prime}{\bf dr^\prime}\prime$ denotes the integral over the filament surface. If the area filling factor of the filaments is $A_{\rm fil}$ 
%then $\int_{S\prime}{dr^\prime} = A_{\rm fil} \int_{S}{\bf dr}$. The last term of Eq.~\ref{eq:ismft} contributes to the excess power and raises 
%the  $\chi^2_{\rm variance} $. We have parameterized  $\chi^2_{\rm variance}$ as a function of $f(\delta_c^2A_{\rm fil})$.

\begin{figure}
\centering
\resizebox{1.0 \hsize }{!}{\includegraphics[angle=0]{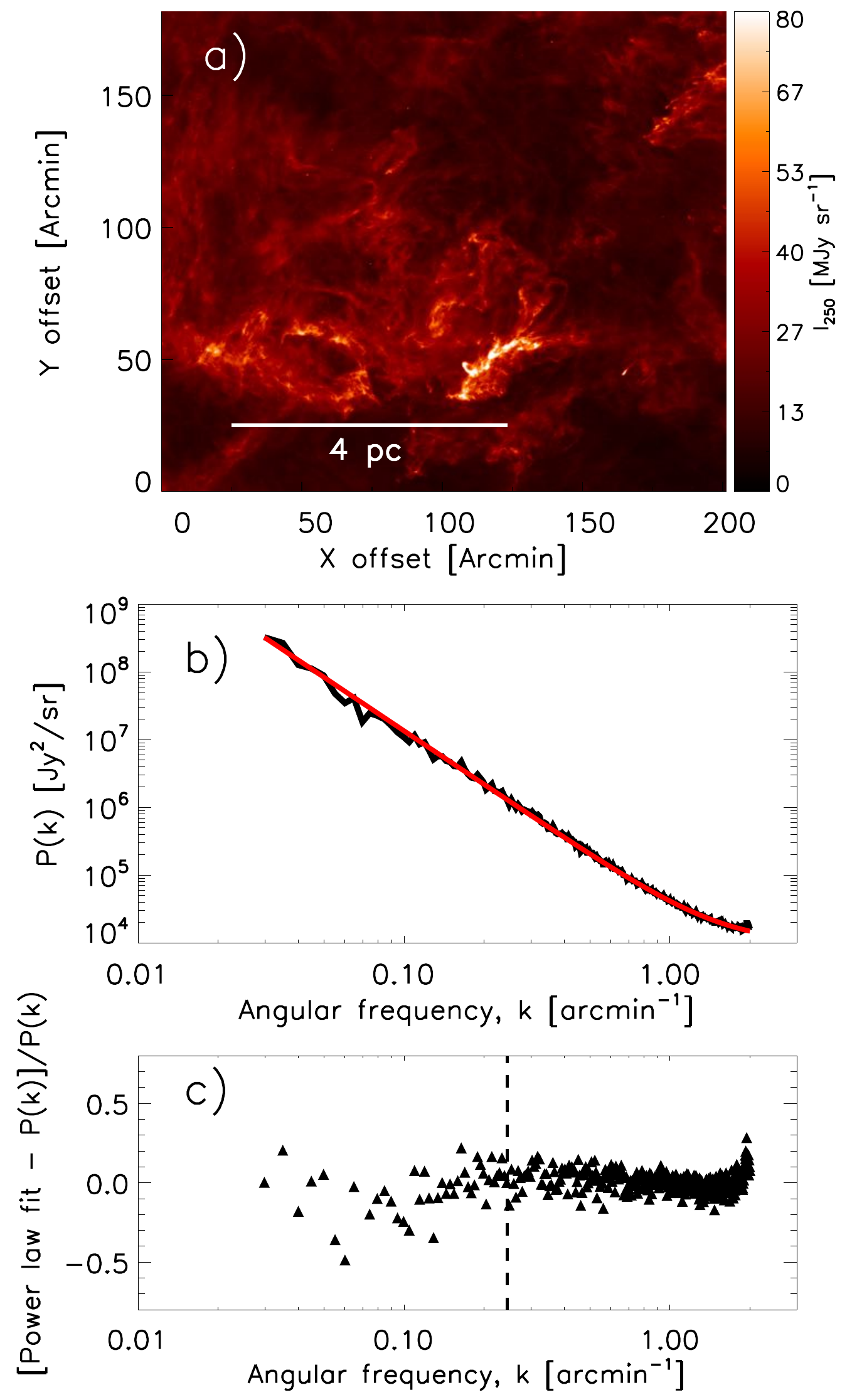}}%{Pipe_fig.ps}
\caption{{\bf a$)$} \her /SPIRE 250 \micron\ emission image of a part of the Polaris cloud. The HPBW angular resolution is 18\farcs2.
 {\bf b$)$} Noise-subtracted and beam-corrected power spectrum of the image shown in panel a$)$ over the range of angular frequencies 0.025 arcmin$^{-1}$ < $k$ < 2  arcmin$^{-1}$.
 The red curve shows  the best fit power-law model over this frequency range, which takes the form $P_{\rm sky}(k)$ =   $A_{\rm ISM}k^{\gamma} $+$ P_0$ 
 with $\gamma$ = -2.63$\pm$0.1. 
 %A constant level of noise was subtracted from the power spectrum and then divided by the power spectrum of
 % \her\ 250~\micron\ beam. 
% In the displayed figure, we show the power spectrum only for the range
%  of angular frequencies, (0.025 arcmin$^{-1}$ < $k$ < 2
%  arcmin$^{-1}$) over which a power-law model, $P_{\rm sky}(k)$ =
 % $A_{\rm ISM}k^{\gamma} $+$ P_0$ was fitted as described in
%  Miville-Desch\^{e}nes et al. (2010).  The dashed magenta curve is
 % the best fit power-law model with $\gamma$ = -2.63$\pm$0.1.  {\bf c$)$} 
    {\bf  c$)$}  Plot of the residuals between the best-fit power-law model and the 
  power spectrum data points 
  %of Polaris image 
  (triangle symbols). The $\chi^2_{\rm Variance}$ (see Eq.~\ref{eq:chi-sq}) of the residuals is $\sim$ 0.03. The
  vertical dashed line marks the angular frequency $k_{\rm fil}\sim (0.6/\theta_{\rm fil})$ corresponding to a 
  characteristic filament width of  $\sim 0.1\,$pc. }
\label{Fig-polaris}
\end{figure}

\begin{figure}
  \centering
  \resizebox{0.88\hsize}{!}{\includegraphics[angle=0]{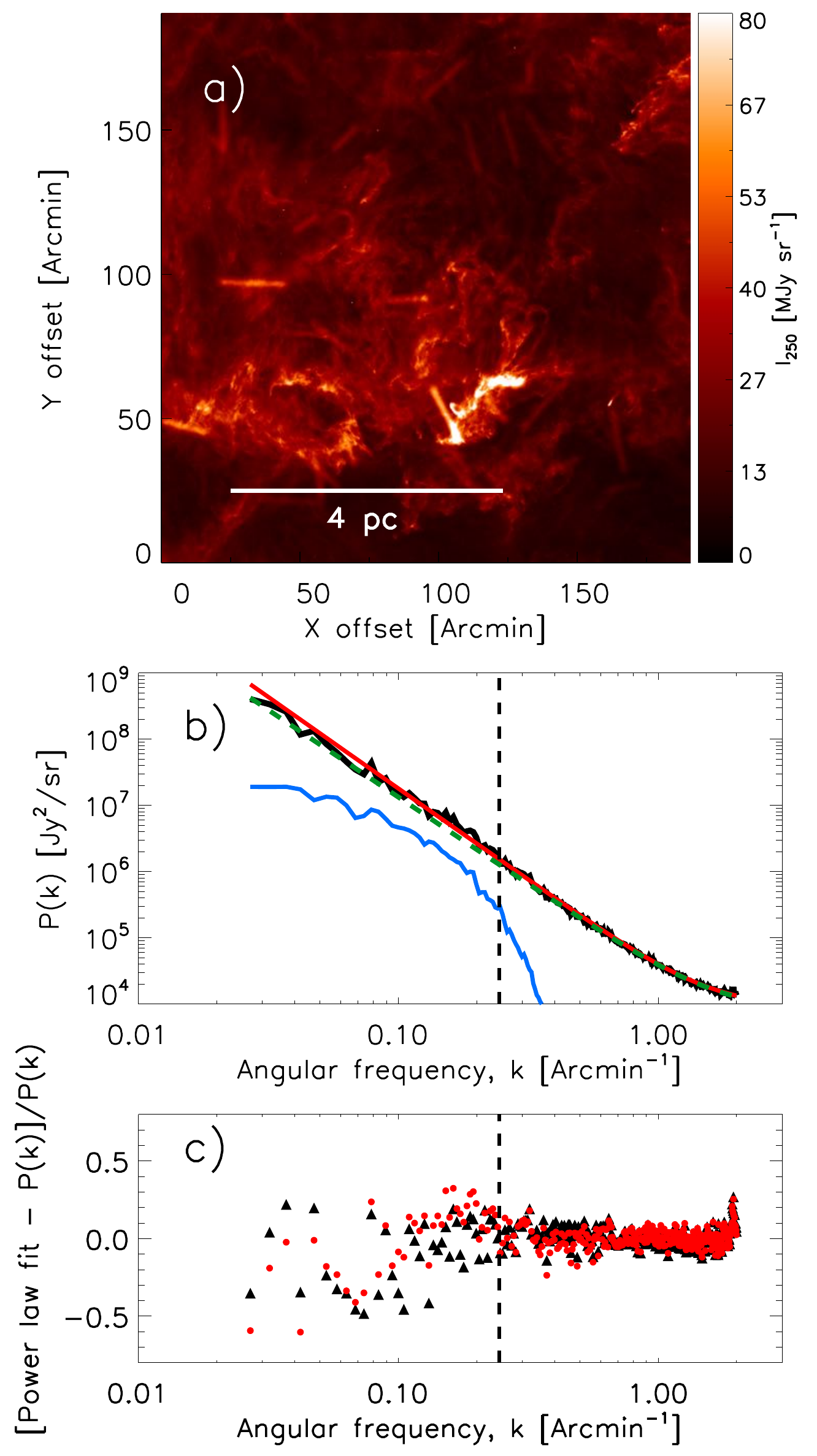}}%{./PolarisImgPowspecBeamNoiseCorrnfil_70_contrast_0.50_DAF_3.67.ps}} %{./Plots/PolarisImgPowspecBeamNoiseCorrnfil_100_contrast_0.50_DAF_5.51.ps}}%{Pipe_fig.ps}}
 %\vspace{0.4cm}
 \caption{Same as Fig.~\ref{Fig-polaris} but for a  250\ \micron\ image with an
 additional population of synthetic filaments. 
 %In this simulation, 
 The population of synthetic filaments has a lognormal distribution of contrasts in the range 0.3 < $\delta_c$ < 2.0 
with a broad peak around $\delta_{\rm peak}\sim$ 0.9.   The overall area filling factor of the synthetic filaments is $A_{\rm fil}$ $\sim$ 3\% and 
 the  $\delta_c^2 A_{\rm fil}$ parameter (see Sect.~\ref{sec:diagnostic}) is  
 %$\delta_c^2 A_{\rm fil}$ = 
 0.023. 
 In {\bf b$)$}, the solid black curve shows the total power spectrum of the original Polaris image plus synthetic filaments. 
 The best-fit power-law (red curve) has  $\gamma$ = $-$2.7$\pm$0.1, 
 %$\gamma$ = $-$2.72$\pm$0.1, 
 slightly steeper than the slope of the
 original power spectrum of the Polaris image shown by the dashed green curve. The blue curve is the power spectrum of
 the image containing only synthetic filaments.  
 In {\bf c$)$}, the black triangles are the same as in Fig.~\ref{Fig-polaris}c and the red dots show the residuals between 
 the best-fit power-law model and the power spectrum of the image including synthetic filaments. 
 The $\chi^2_{\rm variance}$ of the residuals between $k_{\rm min}$ <$k$<1.5 $k_{\rm fil}$  is 0.037. The
 vertical dashed line marks the angular frequency $k_{\rm fil}\sim (0.6/\theta_{\rm fil})$ corresponding to the
 characteristic $\sim 0.1\,$pc width of the synthetic  filaments.
  }
\label{fig:polaris-0.5}
\end{figure}

\begin{figure}
  \centering
  \resizebox{0.88\hsize}{!}{\includegraphics[angle=0]{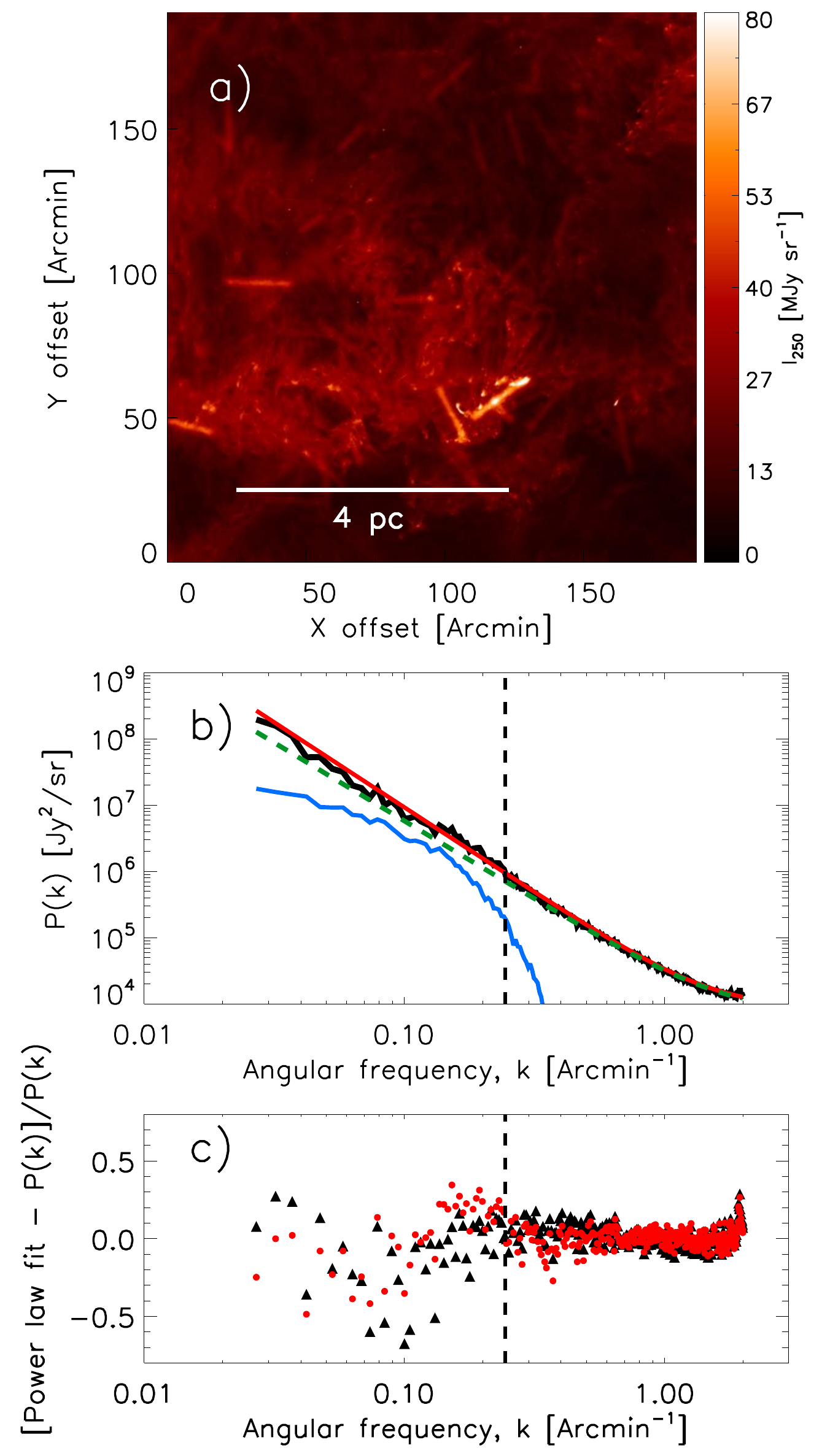}} %{PolarisImgPowspecBeamNoiseCorrWith_filament_nfil_40_contrast_0.50_DAF_2.57.ps}}%{./PolarisImgPowspecBeamNoiseCorrnfil%_70_contrast_0.50_DAF_3.67.ps}} %{./Plots/PolarisImgPowspecBeamNoiseCorrnfil_100_contrast_0.50_DAF_5.51.ps}}%{Pipe_fig.ps}}
%\vspace{0.4cm}
 \caption{ Same as Fig.~\ref{fig:polaris-0.5} for synthetic filaments added to a filament-subtracted image of Polaris obtained with 
\textsl{getfilaments} \citep{Menshchikov2013}.
 %In this simulation, 
 The population of 
 %parameters of the 
 synthetic filaments is the same as that  
 %are  same distribution as 
 in Fig.~\ref{fig:polaris-0.5}. 
  In {\bf b$)$}, the green dashed line shows the best power-law fit to the power spectrum of the filament-subtracted background 
 image of Polaris (with no synthetic filaments), 
% The best power-law fit of the slope is $\gamma $ = $-$2.36$\pm$0.1.
which has a slope of $\gamma $ = $-$2.4$\pm$0.1.
The solid black curve shows the total power spectrum of the filament-subtracted image plus synthetic filaments. 
 The best-fit power-law (red curve) has a slope of $\gamma$ = $-$2.5$\pm$0.1, slightly steeper than the slope 
 of the power spectrum of the filament-subtracted background image (dashed green curve). 
 The blue curve is the power spectrum of the image containing only synthetic filaments.  
 In {\bf c$)$}, the black triangles are the same as in Fig.~\ref{Fig-polaris}c and the red dots show the residuals between 
 the best-fit power-law model and the power spectrum of the image including synthetic filaments. 
 The $\chi^2_{\rm variance}$ of the residuals between $k_{\rm min}$ <$k$<1.5 $k_{\rm fil}$  is 0.034. The
 vertical dashed line marks the angular frequency $k_{\rm fil}\sim (0.6/\theta_{\rm fil})$ corresponding to the
 characteristic $\sim 0.1\,$pc width of the synthetic  filaments.
  }
\label{fig:polaris_filremoved-0.5}
\end{figure}

\section{Power spectrum of the Polaris Herschel data }\label{Sec:polaris}

In this section, we analyze 
the \her/SPIRE image of the Polaris Flare cloud at 250~\micron\ (\citealp{mamd2010,ward-thomson2010}; see also \citealp{schneider2013}), 
%\citep[][see also schneider2013]{mamd2010,ward-thomson2010}, 
which covers an area of 3.0$^{\circ}$$\times $3.3$^{\circ}$ and is shown in 
Fig.~\ref{Fig-polaris}a) at the native (diffraction-limited) beam 
resolution of 18\farcs2. For our analysis, a pixel size of 
6\arcsec\ was adopted\footnote{A zero-level offset of 16.8 MJy/sr was 
also added to the image based on a comparison with {\it Planck} and IRAS 
data.}.  

The Polaris Flare image displays a spectacular distribution of low column
density filaments. All of these filaments are thermally subcritical.
The mean peak surface brightness contrast of these filaments over the 
%compared to the 
local background is around $<\delta_c> \sim$ 0.9, but the filaments occupy 
only a small fraction $\sim 2\% $ of the total surface area, leading to $\delta_c^2 A_{\rm fil}\sim$ 0.016. 
%very low area filling factor of 0.02\%, amounting $\delta_c f_{\rm fil}\sim$ 0.024. 
%
%Previous observations have shown that for low column
%density filaments in a non-star forming cloud ( e.g., filaments in the
%Polaris flare translucent cloud) the average peak filament contrast,
%$\delta_c = (I^{\rm peak} - I^{\rm bkg})/I^{\rm bkg}$, with respect to
%the background was around $\sim$ 1 but with extremely low area  (Arzoumanian et al. 2017, {\it in
%  prep}).

To first order, the transverse structure of the Polaris filaments
is well described by Gaussian profiles with a FWHM of $\sim$ 0.1 pc
(assuming a distance $\sim$140 pc for the Polaris cloud) \citep[][]{arzoumanian2011,arzoumanian2017}.  
%The total
%molecular H$_2$-gas mass content is estimated to be around $\sim 2
%\times$10$^3$ \msol, of which only $x$\% of the gas mass resides along
%the prominent filaments.  
\cite{mamd2010}  carried out a power spectrum analysis for the Polaris image
over spatial scales ranging from 0.01 pc to 10 pc.
%studied the statistical properties of structures in the Polaris Flare ranging from
%spatial scales of 0.01 pc to 10 pc using a power spectrum analysis. 
The power spectrum revealed a continuous power-law, $P(k)\propto k^{\gamma}$, with an exponent of
$\gamma =- 2.65$ down to the scale of the beam, suggesting a scale-free image. 
%a reminiscent of scale-free structures. 
Following the same scheme as Miville-Desch\^{e}nes et
al. (2010), we derived the power spectrum of a sub-field\footnote{In
order to capture the maximum rectangular area within the Polaris image for
easier computation of the power spectrum, we rotated the original map in equatorial coordinate by 
13.6$^{\circ}$ clockwise about its center and extracted the largest area
excluding turn-around data points near the edges of the field.}  of Polaris
shown in Fig.~\ref{Fig-polaris}. Prior to computing of power spectrum, we apodized the edges of the image 
by a sine function to ensure smooth periodic boundary condition.  After subtracting the noise power
spectrum level estimated from the mean power at angular frequencies
$k>$ 3.5 arcmin$^{-1}$, we corrected for convolution effects by dividing the observed power spectrum
by the power spectrum of the {\it Herschel} telescope beam at 250~\micron\  \citep{martin2010} obtained from scans of Neptune. 
In order to derive the power spectrum slope, we then fitted a
power-law model of the form $P_{\rm sky}(k) = A_{\rm ISM}k^{\gamma} + P_0 $
to the corrected power spectrum over the range of angular frequencies $0.025 ~{\rm  arcmin^{-1}}~<k<~ 2$~arcmin$^{-1}$, as described in \cite{mamd2010}.
In Fig.~\ref{Fig-polaris}b, we show the power spectrum of the Polaris image 
over the range of angular frequencies used in the power-law fit.  
The 
%dashed magenta 
red curve represents the best-fit power law 
%best-fit model power spectrum 
with $\gamma$ =$-$2.63$\pm$0.1, which is very close to the $\gamma$
=$-$2.65$\pm$0.1 value obtained by Miville-Desch\^{e}nes et al. (2010).
Visual inspection shows that there is no clear spectral
signature of a characteristic scale embedded in the observed power spectrum. 
Figure~\ref{Fig-polaris}c shows the residuals [$P_{\rm Best-fit}(k) - P_{\rm Polaris}(k)$]/$P_{\rm Polaris}(k)$ as a function of angular
frequency. 
%The plot of residuals should in principle capture 
Any significant  kink or distortion in the power spectrum due to the presence of a characteristic scale 
%the characteristic inner width of filaments 
should in principle be captured as a significant deviation from zero in the plot of residuals.
The plot of residuals for the Polaris Flare image (Fig.~\ref{Fig-polaris}c) 
does not exhibit such a deviation. 

In order to critically analyze this finding, we performed a suite of numerical experiments
by injecting synthetic filaments separately into 1) the original {\it Herschel}/SPIRE 250~\micron\ image of Polaris,   
 and 2) the filament-subtracted image obtained after applying the \textsl{getfilaments} algorithm \citep{Menshchikov2013}
to the {\it Herschel}/SPIRE image to remove most of the real filamentary structures. 
%with the \textsl{getfilaments} algorithm \citep{Menshchikov2013}. using {\it Getfilament} algorithm. 
We repeated the same power spectrum analysis as described above on both sets of modified {\it Herschel} images 
(see Sects.~\ref{subsec:polaris-syn} and \ref{Sect:polaris-bkg} below). 
For this analysis, we preferred to use SPIRE  250~\micron\ data rather than 18\farcs2 column density images 
produced from the combination of {\it Herschel} data at 160~\micron\ to 500~\micron\  \citep[cf.][]{palmeirim2013} 
%(an HGBS product, \citealp{palmeirim2013}) 
because the former are less affected by noise and better behaved 
from a power-spectrum point of view \citep[cf.][]{mamd2010}.

%\subsection{Construction of a synthetic image with low column density  filaments} \label{subsec:polaris-img}
\subsection{Construction of an image with synthetic  filaments} \label{subsec:polaris-img}

To create a synthetic filament image, the first step was to generate a
map of randomly oriented 1D delta line functions as described in Sect.~\ref{sec:single-filament}. 
Then, we convolved this initial synthetic map with a Gaussian kernel such that the
projected spatial FWHM width of the kernel was 0.1 pc as described in
Sect~\ref{sec:single-filament}. 
When creating synthetic filaments we neglected the fluctuations observed along real {\it Herschel} filaments \citep{roy2015},  
because the contrast of these fluctuations above the average filament 
%column density 
is $<<1$, and also the area filling factor 
of these fluctuations is very small.  To maximize the effect of
a characteristic width in our simulations, we fixed the FWHM width of
the Gaussian filaments to a strictly constant value.  
We controlled the contrast parameter $C$ of each filament by 
%with respect to
measuring the local background emission in the close vicinity of the filament within the background image.   The distribution of the contrast parameter was chosen to reflect the observed distribution in each region. 
In the simulation, we varied the angular length of the filaments randomly between a
minimum of 30$\times$18\farcs2 =546\arcsec\ and a maximum of 70$\times$18\farcs2 = 1274\arcsec, corresponding to 0.4 to 0.9 pc at $d=140$ pc. 
Figure ~\ref{fig:polaris-0.5}a shows one such realization including a population of synthetic
filaments with a lognormal distribution of contrasts in the range $0.3 < \delta_c < 2.0$, 
co-added to the original map of Polaris. In this example, the population of synthetic filaments
has an area filling factor $A_{\rm fil}\sim 3.2\% $. 
%$A_{\rm fil}\sim 7.3\% $. 
%The population of synthetic filaments has a lognormal distribution of contrasts in the range 0.3 < $\delta_c$ < 2.0 
%with a peak $\delta_{\rm peak}\sim$ 0.6.

\subsection{Effect of synthetic filaments on the power spectrum of the Polaris image}
 \label{subsec:polaris-syn}

Next, we investigated the effect of synthetic filaments on the power spectrum of the Polaris original image on one hand,  
and the power spectrum of the filament-subtracted Polaris image on the other hand.
First, we discuss  the case of the Polaris  original image.

Figure~\ref{fig:polaris-0.5}b shows the total power spectrum of the Polaris image in Fig.~\ref{fig:polaris-0.5}a, 
which includes a population of synthetic filaments with a log-normal distribution of contrasts, $\delta_c$.  
In Fig.~\ref{fig:polaris-0.5}, the range of  contrast values  in the synthetic distribution varied 
in the range 0.3 < $\delta_c$ < 2.0 
%between 0.2 $< \delta_c$ 2 
with a peak at $\delta_{\rm peak} \sim 0.9$. The weighted average of the contrast over the length of the simulated filaments is <$\delta_c$> $\sim$ 0.85.   
%contrast level $\delta_c \sim $0.5. 
In Fig.~\ref{fig:polaris-0.5}a, the synthetic filaments are clearly visible against their local background.  
The best power-law fit to the power spectrum (red curve) has 
a logarithmic slope {\bf $\gamma$=$-$2.7$\pm$0.1}, slightly steeper than the slope
of the Polaris original  image. 
The vertical dashed line marks the angular frequency, $k_{\rm fil}$ = $\Gamma$ $\sim (0.6/\theta_{\rm  fil})$
%$\footnote{The power spectrum of a Gaussian beam is also a
%  Gaussian, and its $FWHM$ $\Gamma$ is related to the $FWHM$ width
%  $\theta_{\rm fil}$ through relation $\Gamma =
%  \sqrt(8)ln(2)/\pi\theta_{\rm fil}$ $\sim$ 0.6/$\theta_{\rm fil}$.}
%
$ \sim 0.24$ arcmin$^{-1}$, corresponding to the characteristic angular width of the synthetic filaments, 
$\theta_{\rm fil} =$ 147\arcsec\ (i.e., 0.1 pc at $d$~=~140 pc). 
Comparison of Fig.~\ref{fig:polaris-0.5}b to Fig.~\ref{Fig-polaris}b shows that the 
%inclusion of 
synthetic filaments contribute 
an insignificant amount of power around $k_{\rm fil} =0.24$ arcmin$^{-1}$, 
which can hardly be detected without prior knowledge of the power spectrum of the original ISM image.  
Figure~\ref{fig:polaris-0.5}c plots the normalized residuals between the
best power-law fit and the power spectrum data, $[P_{\rm Best-fit}(k) - P_{\rm Polaris}(k)]/P_{\rm Polaris}(k)$, as a function of angular frequency. 
These residuals (red filled circles) can be compared with the residuals obtained with the original image, represented by black 
triangles in both Fig.~\ref{Fig-polaris}c and Fig.~\ref{fig:polaris-0.5}c. 
Again, no clear signature of the presence of synthetic filaments can be detected despite the fact that they have a characteristic width.

Now let us investigate the power spectrum of each component more
closely to understand the absence of any detectable signature in the total power spectrum.
The blue curve and the green dashed curve in Fig.~\ref{fig:polaris-0.5}b 
show the power spectrum of the synthetic filament image and that of the original 
image, respectively.  
%Since the randomly inserted synthetic filaments
%are not correlated with the diffuse structures of the original Polaris
%image, the power spectrum of synthetic filament plus original Polaris
%image is equivalent to the sum of $P_{\rm fil}(k) +P_{\rm
%  Polaris}(k)$. 
Note that the power spectrum of the filament image,
$P_{\rm fil}(k)$, is lower than the power spectrum  $P_{\rm Polaris}(k)$ at
$k=k_{\rm fil}$ by a factor of $\sim$ 5. 
This is because the population of synthetic filaments only have moderate area filling factor ({\bf $A_{\rm fil} \sim 3.2\%$}) 
and contrast ({\bf <$\delta_c$>~$\sim 0.85$ }). 
In this experiment, the product of the area filling factor with the  square of the column density contrast ($A_{\rm fil}\delta^2_c\sim$~0.02) 
was in agreement with the real filaments observed in Polaris 
(which have $A_{\rm fil}\,$<$\delta_c$>$^2 \sim 0.01$ -- cf. \citealp{arzoumanian2017}). %(cf. \citealp{mamd2010,arzoumanian2017}).
The additional power introduced by the synthetic
filaments is not localized in the vicinity of $k_{\rm fil}$ but rather
spread out at angular frequencies $k \la k_{\rm  fil}$, following 
a shallow power-law, 
whereas  the power at high angular frequencies at $k> k_{\rm fil}$ drops sharply. 
Given the choice of $\delta_c$ and $A_{\rm fil}$ made here, the  relative contribution of filaments to the
total power spectrum, $P_{\rm fil}(k)/P_{\rm Polaris}(k)$ is highest
in the vicinity of $ k_{\rm fil}$, but too small to create any
detectable feature in the power spectrum.

When the contrast and/or filling factor of the synthetic filaments is gradually increased, the spectral
imprint in the resulting power spectrum becomes more and more pronounced.  
Figure~\ref{fig-Polaris-extreme}a in Appendix~\ref{Appen:extreme} shows a simulated  
image including a population of synthetic 0.1-pc filaments with 
contrast {\bf $\delta_c\sim 1.1$} and area filling factor {\bf $A_{\rm fil}\sim 7.2\%$}.
This is quite an extreme scenario for a non-star-forming molecular cloud with low column density such as Polaris. 
Figure~\ref{fig-Polaris-extreme}b shows the corresponding power spectra, 
which should be compared to those in Fig.~\ref{fig:polaris-0.5}b. 
%
%The solid black curve in Figure \ref{fig-Polaris-extreme}b) shows the
%total power distribution as a function of angular frequency. The solid
%blue curve denotes the level of power as a function of angular
%frequency contributed by the population of high contrast synthetic
%filaments.  
%
%Compared to the power spectrum with contrast level of 0.5
%(see Fig.~\ref{fig:polaris-0.5}b in Appendix \ref{Appen:extreme}),
It can be seen that the amplitude of the synthetic power spectrum in Appendix~\ref{Appen:extreme} 
(with contrast $<\delta_c\>\sim 1.1 $ and $A_{\rm fil} \sim$ 7.2\%)
is higher by a factor of 3 to 4 than the amplitude of the power spectrum of Fig.~\ref{fig:polaris-0.5}b
( with contrast  <$\delta_c$>~$\sim 0.85$ and $A_{\rm fil} \sim$ 3.2\%),  mostly due to the increase in the combination 
of contrast parameter and area-filling factor <$\delta_c$>$^2 A_{\rm fil}$  
%$<\delta_c>^2 A_{\rm fil}$} 
between the two simulations{\bf  [$\sim (1.1/0.85)^2 \times (7.2 / 3.2) ( \sim 3.7 $]}.
%\st{(Note that the area filling factors in both simulations are similar, $A_{\rm fil}\simeq 6\%$.) } 
The red curve in Fig.~\ref{fig-Polaris-extreme}b
shows the best power-law fit which has a logarithmic slope $\gamma =$ $-$2.96$\pm$0.1. 
At $k \la k_{\rm fil}$, there is a significant enhancement of power due to the fact the 
synthetic filament power spectrum $P_{\rm fil}(k)$ is now comparable to 
the Polaris power spectrum $P_{\rm Polaris}(k)$ at $k \la k_{\rm fil}$.  
Accordingly, in this case, the residuals between the best power-law fit and the total power spectrum data 
depart significantly from zero at $k \la k_{\rm fil}$ (cf. Fig.~\ref{fig-Polaris-extreme}c).
It is to be borne in mind, however, that the population of synthetic filaments used in Appendix~B
have much higher contrast and area filling factor than the actual filaments of the Polaris cloud 
(compare Fig.~\ref{fig-Polaris-extreme}a and Fig.~\ref{Fig-polaris}a).
 
 \subsection{Effect of synthetic filaments on the power spectrum of the filament-subtracted image} \label{Sect:polaris-bkg}
 
 So far we have explored the response of the power spectrum to a synthetic population of filaments injected 
 into the original {\it Herschel} image, which itself includes emission from real filamentary structures. 
 It is instructive to assess the extent to which the real filaments 
 %whether the real filaments 
 %which are already 
 present in the image may reduce the relative contribution of synthetic filaments. In order to evaluate this 
 we adopted two approaches -- first, we subtracted the emission of at least the most prominent real filamentary structures 
 from the Polaris 250~\micron\ image using the \textsl{getfilaments} algorithm \citep[][see Fig.~\ref{fig-simPolaris}b in Appendix~B 
 for the resulting filament-subtracted image]{Menshchikov2013} 
 and then repeated the same experiment as described in Sect.~\ref{subsec:polaris-img}. 
 Second, we examined the effect of filaments embedded in a typical scale-free synthetic cirrus images 
 (See Appendix~\ref{synthetic-bkg}).
 In order to be consistent, we used the same population of synthetic filaments as 
 in Sects.~\ref{subsec:polaris-img} and \ref{subsec:polaris-syn}.
 
Figure~\ref{fig:polaris_filremoved-0.5} summarizes the effect of the synthetic $0.1\,$pc filaments 
on a background image which is essentially devoid of real filamentary structures. 
Although the logarithmic  power-spectrum slope of the background image is shallower ($\gamma_{\rm bkg} = -2.5$) 
than that of the Polaris original image ($\gamma_{\rm obs} = -2.7$), 
the overall morphology remains the same. 
In particular, the power spectrum of the synthetic filament component is still significantly lower than 
the total power spectrum of the background image, even though the power
% despite the subtraction of the power 
arising from real filaments has been subtracted from that image. 
%removing the observed filaments. 
The $\chi^2_{\rm variance} $ of the filament-subtracted background image is 0.04, very close 
the  $\chi^2_{\rm variance}$ value obtained for the Polaris original image. 
Moreover, there is still no clear signature of the presence of synthetic filaments 
in the residuals plot  (Fig.~\ref{fig:polaris_filremoved-0.5}c).  
Similar conclusions were reached in Appendix~\ref{synthetic-bkg} in the case of synthetic filaments added
to a purely synthetic background image.

\begin{figure} 
 \resizebox{.95\hsize}{!}{\includegraphics[angle=0]{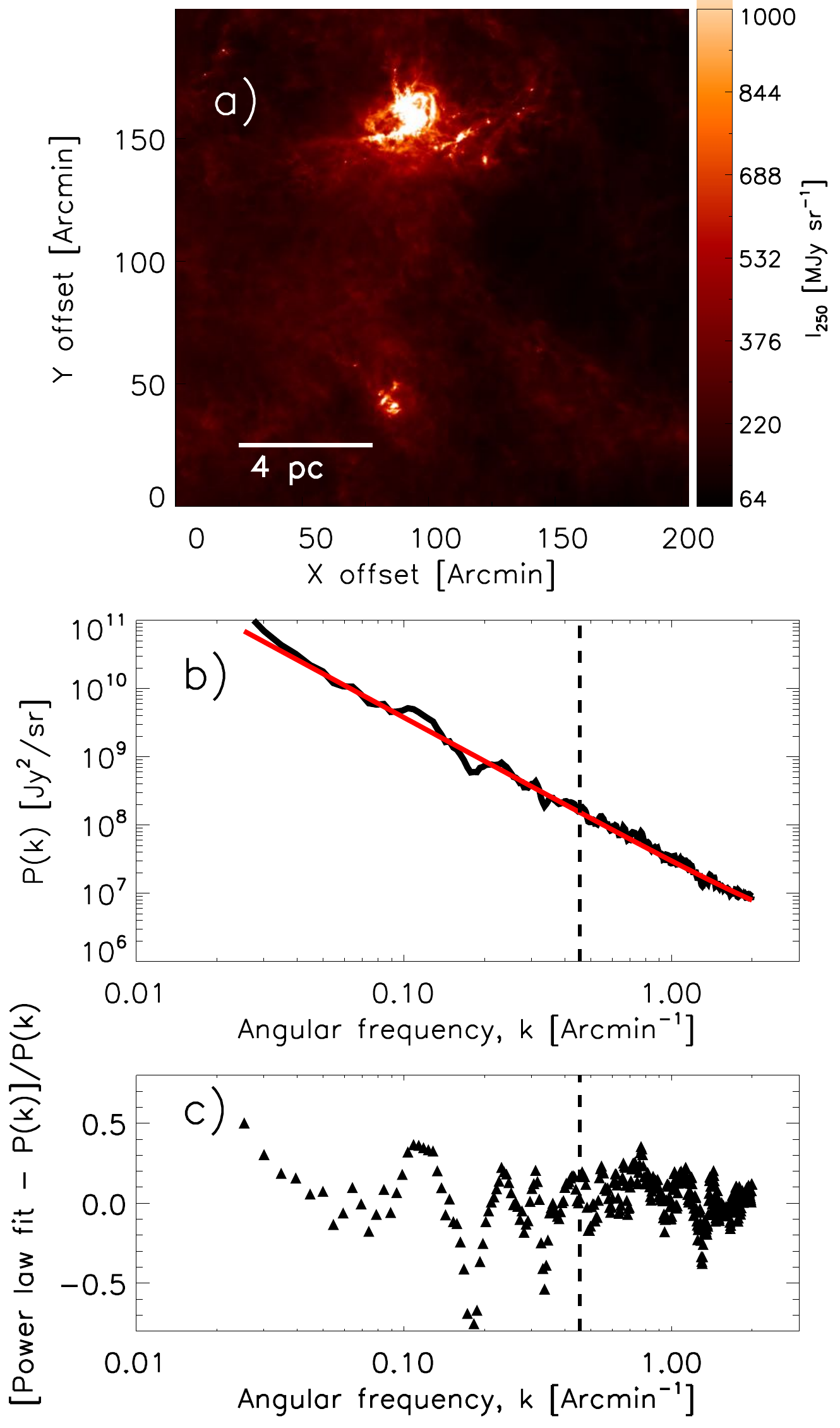}}%%nfil_100_contrast_0.00_.ps}}%Aquila_and_filament_powspecnfil_100_contrast_0.00_.ps}}
 \caption{{\bf a$)$} \her /SPIRE 250 \micron\ image of the Aquila cloud at the native resolution of 18\farcs2. 
 %but rotated in equatorial coordinates by 31.4$^\circ$ in clockwise direction.
 {\bf b) } Noise-subtracted and beam-corrected power spectrum of the image shown in panel a$)$ over the range of angular frequencies 0.025 arcmin$^{-1}$ < $k$ < 2  arcmin$^{-1}$ 
 (black curve).  The red curve shows  the best fit power-law model over this frequency range, which has a logarithmic slope $\gamma = -2.26\pm 0.1$. 
 %takes the form $P_{\rm sky}(k)$ =   $A_{\rm ISM}k^{\gamma} $+$ P_0$ 
 The vertical dashed line marks the angular frequency  $k_{\rm fil} \sim (0.6/\theta_{\rm fil-width}) \sim 0.45\,$arcmin$^{-1}$ corresponding to a filament width 
 of $\theta_{\rm fil-width}$ =79\arcsec\ (FWHM), i.e., 0.1 pc at a distance of 260 pc.
  {\bf c) } Plot of the residuals between the best power-law fit and the 
  power spectrum data points  (triangle symbols). The $\chi^2_{\rm Variance}$ of the residuals between $k_{\rm min}$ <$k$<1.5 $k_{\rm fil}$ is $\sim$ 0.045 }
\label{fig:Aquila-img}
\end{figure}

\begin{figure}
  \resizebox{\hsize}{!}{\includegraphics[angle=0]{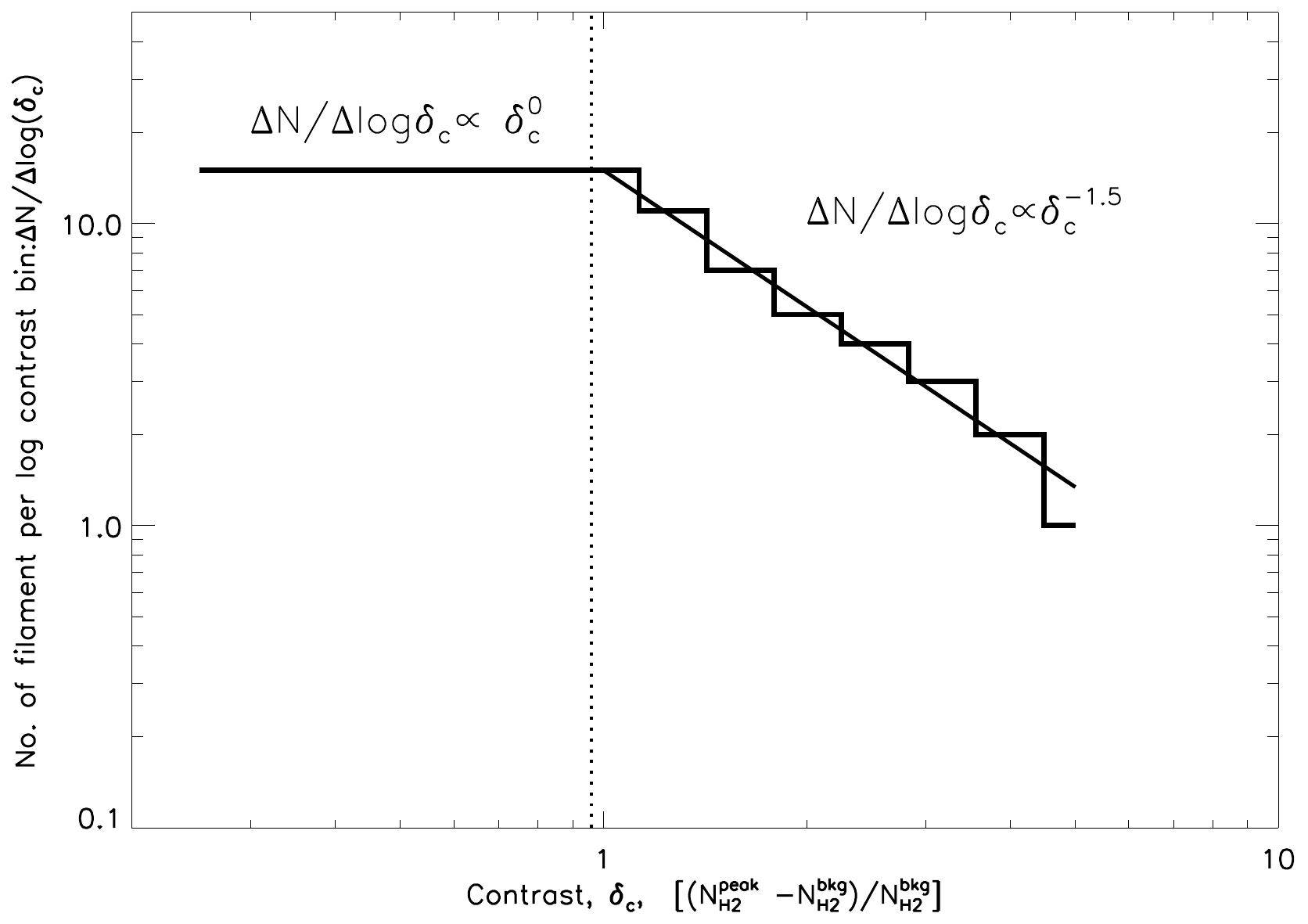}}%delta_nfil_100_delta_mmp0.20_3.00_1.00DAF_6.71.ps}}%{Pipe_fig.ps}}
\vspace{0.1cm}
 \caption{Two-segment power-law approximation (black solid lines)  to the distribution of  
   filament column density contrasts observed in the Aquila molecular cloud  \citep{arzoumanian2017}:
    d$N/{\rm dlog}(\delta_c)$ $\sim$ $const $ for 0.3 $\leq
   \delta_c \leq$ 1, and d$N/{\rm dlog}(\delta_c)$ $\sim$ $\delta_c^{-1.5} $
   for 1 $\leq \delta_c \leq 4$. The vertical dotted line marks the average filament contrast $<\delta_{\rm c}>\ \sim 1$. 
%   The peak of the contrasts distribution is pivoted at $\delta_{\rm peak}$ =1.  
 The overplotted histogram shows the distribution of column density contrasts for the population of 100 synthetic filaments 
 used in the simulation of Sect.~4.
   }
\label{fig-contrast-dist}
\end{figure}

\begin{figure}
\centering
\resizebox{0.95 \hsize }{!}{\includegraphics[angle=0]{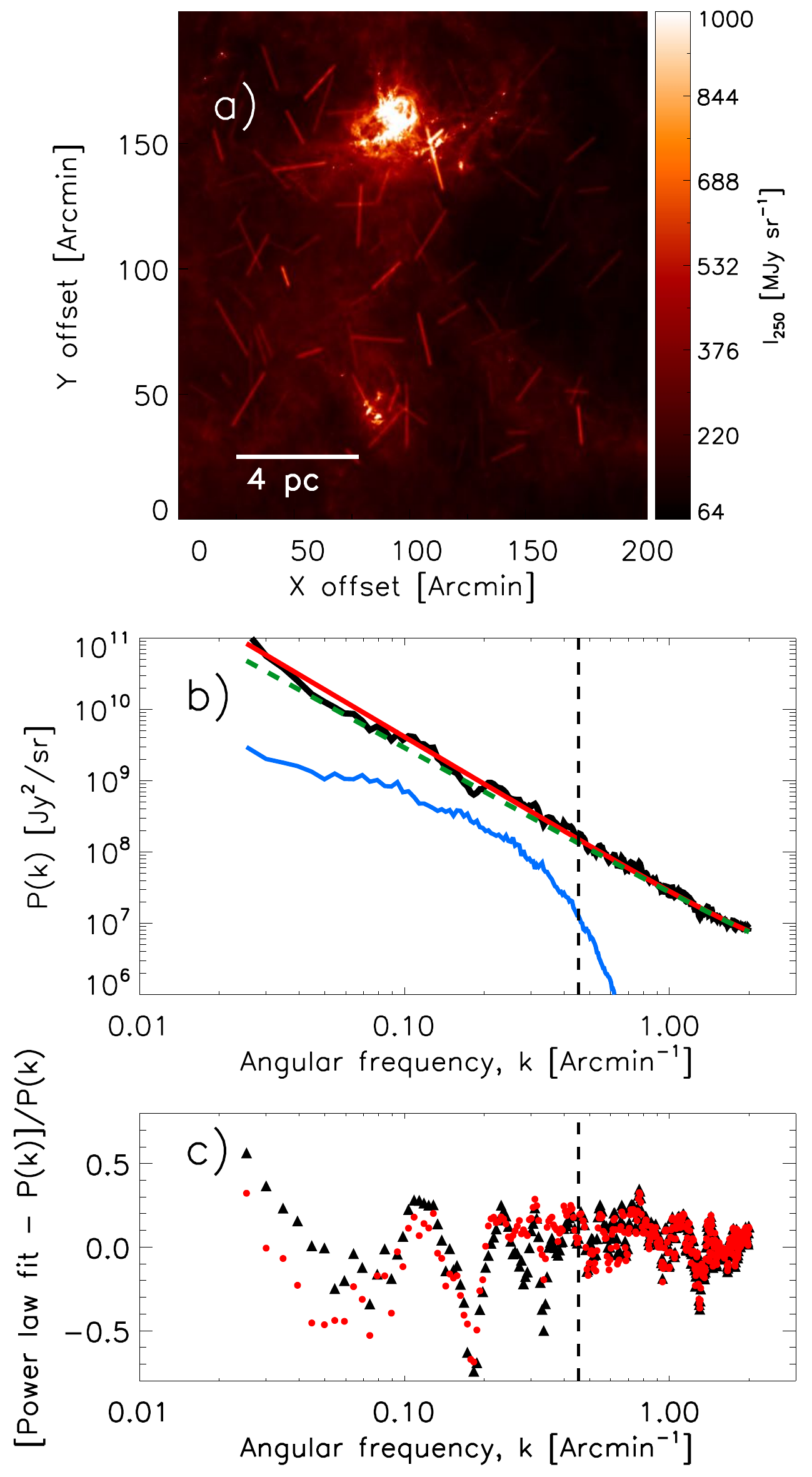}}
%AquilaImgPowspecBeamNoiseCorrnfil_100_delta_mmp0.30_4.00_1.00DAF_5.65.ps}}
%{./AquilaImgPowspecBeamNoiseCorrnfil_100_delta_mmp0.30_3.00_1.00DAF_5.71.ps}}%AquilaImgPowspecBeamNoiseCorrnfil_100_delta_mmp0.30_3.00_1.00DAF_5.97.ps}}%{Pipe_fig.ps}
\caption{Same as Fig.~\ref{fig:Aquila-img} but for a simulated image including a population of synthetic filaments
with a realistic distribution of column density contrasts (see Fig.~\ref{fig-contrast-dist}) 
added to a filament-subtracted image of the Aquila cloud.  
{\bf a$)$} Simulated image. 
  The  weighted average contrast $<\delta_c>$ of the distribution of synthetic filaments is
  0.96 and the total area covering factor $A_{\rm fil}$ is 5.5\%, leading to $\delta_c^2A_{\rm fil}\sim0.051$.    
  {\bf b$)$}  Power spectrum of the simulated image (black solid curve).
   The red curve corresponds to the best power-law fit with $\gamma =$ $-$2.3$\pm$0.1. 
   For comparison, the best power-law fit to the power spectrum of the Aquila original image is
   over-plotted as a green dashed line. 
{\bf c$)$} Residuals between the best power-law fit and the 
  power spectrum of the simulated image (red solid circles). The $\chi^2_{\rm Variance}$ of the residuals is $\sim 0.054$
  For comparison, the black filled triangles  show similar residuals for the Aquila original image 
  (cf. Fig.~\ref{fig:Aquila-img}c).}
\label{Fig-Aquila-bkg}
\end{figure} 
 
%\section{Exploring the parameter space with more realistic simulations in the Aquila cloud} \label{Sec:Aquila}
\section{Exploring the parameter space with 
%more realistic 
simulations in the Aquila cloud} \label{Sec:Aquila}

%So far we have explored the power spectrum response of a population of
%filaments with constant contrast ($\delta_c$) and width.  In actual
%observations, however, there is a distribution of these physical
%parameters. It is instructive to explore the parameter space with a
%more realistic population of synthetic filaments. 
%resembling the true sky.
%Fortunately, 
The Aquila molecular cloud harbors a statistically
significant number of filaments with a wide range of filament column
density contrasts \citep{konyves2015, arzoumanian2017}, allowing us to derive a realistic distribution of
contrasts which can then be used for constructing more realistic populations
of synthetic filaments.
\subsection{Observed filament properties in Aquila}

In contrast to the Polaris cloud, the Aquila molecular cloud is an
active star forming complex at a distance\footnote{The distance of the Aquila cloud is uncertain, with values ranging 
from 260 pc to 414 pc in the literature. Assuming the upper 
distance value would push $k_{\rm fil}$ toward higher angular frequencies in Fig.~\ref{fig:Aquila-img}b,c,  
making the detection of the 0.1 pc scale even more difficult in the power spectrum.} 
of 260 pc, including several supercritical
filaments \citep{andre2010, konyves2015}. 
Figure~\ref{fig:Aquila-img}a shows the {\it Herschel}/SPIRE 250~\micron\ image 
of the Aquila cloud, which covers a projected sky area of 3.4$^{\circ}$ $\times$ 3.2$^{\circ}$.
The corresponding power spectrum is shown in Fig.~\ref{fig:Aquila-img}b.
%Figure~\ref{fig:Aquila-img}b shows a noise subtracted and convolution-corrected
%power spectrum of the image shown in Fig.~\ref{fig:Aquila-img}a over
%the angular frequency range 0.025 arcmin$^{-1}$ $< k <$ 2
%arcmin$^{-1}$.  The over-plotted red curve shows the best-fit model
%power spectrum which has a slope of $\gamma$ = $-$2.1$\pm$0.12.  The
%vertical dashed line corresponds to the angular scale $k_{\rm fil}
%\sim$ 0.45 arcmin$^{-1}$ for a 0.1 pc scale at a
%distance of 260 pc.
%

As part of a systematic analysis of filament properties in nearby clouds based on HGBS data,  
\citet{arzoumanian2017} took a census of filamentary structures in Aquila. They obtained 
a distribution of filament column density contrasts which can be conveniently approximated 
by the two-segment power law shown in Fig.~\ref{fig-contrast-dist}: 
d$N/{\rm dlog}(\delta_c)$ $\sim$ $const $ for 0.3 $\leq \delta_c
\leq$ 1, and d$N/{\rm dlog}(\delta_c)$ $\sim$ $\delta_c^{-1.5} $ for 1 $\leq
\delta_c \leq 4$. 
This observed distribution of filaments contrasts has a peak around $\delta_c^{\rm peak}~\sim 1$ 
and spans a  broad range from 
low $\delta_c \sim 0.3$ values  to fairly high $\delta_c \sim 4$ values.
%in the Aquila molecular cloud 
%has a high dynamic range. Toward the low contrast regime, a few
%filaments are detected with a contrast as low as $\delta_c$ $\sim$ 0.2, whereas
%on the higher contrast side a few thermally supercritical filaments
%are detected with $\delta_c$ as high as $\sim$ 3 (Arzoumanian et
%al. 2017, also see Fig.~\ref{fig-contrast-dist}).  
%The distribution has a peak around $\delta_c^{\rm peak}~\sim$ 1. The overall distribution of 
%column density contrasts observed in the Aquila region can be
%approximated by a two segment power-law distributions:
%$dN/dlog(\delta_c)$ $\sim$ $\delta_c^{1.67} $ for 0.2 $\leq \delta_c
%\leq$ 1, and $dN/dlog(\delta_c)$ $\sim$ $\delta_c^{-1.5} $ for 1 $\leq \delta_c \leq 3$.  
%
%Figure~\ref{fig-contrast-dist} shows the resulting contrast distribution for 100 synthetic filaments. 
The weighted average column density contrast of the filaments observed in Aquila is  $<\delta_c>~\sim 1$, 
and their area filling factor is $A_{\rm fil}~\sim$ 3\%.  
While the census of filaments obtained by \citet{arzoumanian2017} may be affected by incompleteness 
issues for low-contrast\footnote{Given the fact that the amplitude of a power spectrum $\propto \delta_{c}^2$,  undetected filaments (with low contrasts) below the completeness level will not have any significant effect on the net amplitude of synthetic filaments power spectrum.} ($\delta_c << 1$)  filaments, it should be essentially complete for high-contrast  ($\delta_c \ga 1$) 
supercritical filaments.

%\subsection{Power spectrum analysis for the Aquila field}
\subsection{Effect of a synthetic population of filaments on the power spectrum}

Using a methodology similar to that employed in Sect.~\ref{Sec:polaris} 
%Sect.~\ref{subsec:polaris-img} 
for Polaris, we added a population of synthetic filaments with fixed 0.1~pc width 
to a filament-subtracted {\it Herschel} image of the Aquila region at 250 \micron . 
%the {\it Herschel} 250 \micron\ image of the Aquila cloud. 
The distribution of column density contrasts\footnote{The column density contrasts of cold molecular filaments are somewhat higher than their surface brightness contrasts at $250\, \mu$m. 
To be on the conservative side, we used the observed distribution of column density contrasts for constructing synthetic filaments in the $250\, \mu$m images. 
The actual surface brightness contrasts are actually lower than what we assumed here.} 
for the synthetic filaments was constructed to be consistent with observations and is represented by the histogram 
in Fig.~\ref{fig-contrast-dist}. The weighted mean contrast of the whole population of
synthetic filaments was <$\delta_{\rm c}$>~$ \sim 0.96$. 
 Like in the Polaris case, the background image was obtained from the {\it Herschel}/SPIRE $250\, \mu$m 
of the Aquila cloud
image after removing observed filaments using the \textsl{getfilaments} algorithm \citep{Menshchikov2013}. 
The resulting synthetic image is shown in Fig.~\ref{Fig-Aquila-bkg}a.
%Fig.~\ref{Fig-Aquila}a.  
%Figure \ref{Fig-Aquila}a) shows the image of Aquila molecular cloud
%coadded with a population of synthetic filaments with a controlled
%distribution of parameters,
%replicating a realistic contrast distribution consistent with
%observations (see Fig.~\ref{fig-contrast-dist}).  The weighted mean contrast of this population of
%filaments is $<\delta_{\rm c}>\sim $ ~ 1.08.  
%Figure~\ref{Fig-Aquila}b 
Figure~\ref{Fig-Aquila-bkg}b 
shows the power spectrum of each component in the synthetic image: 
the blue curve corresponds to the contribution of the synthetic filament distribution, while 
the black curve is the total 
%noise-subtracted and  beam-corrected 
power spectrum of the Aquila background plus filament image.  
%The red line in Fig.~\ref{Fig-Aquila}b is the best-fit power law to
%the total power spectrum, [$P(k)_{\rm Aquila} + P(k)_{\rm fil}$], which has a logarithmic slope 
%$\gamma=-2.3\pm 0.1$,  only slightly steeper than the slope of
%the original power spectrum (green dashed line).
%The dashed green line is the best power-law fit ($\gamma = -2.1\pm0.1$)
%to the 
%noise-subtracted and beam-corrected 
%power spectrum of original
%Aquila image.

\begin{figure*}
%  \centering
  \resizebox{ 0.49\hsize}{!}{\includegraphics[angle=0]{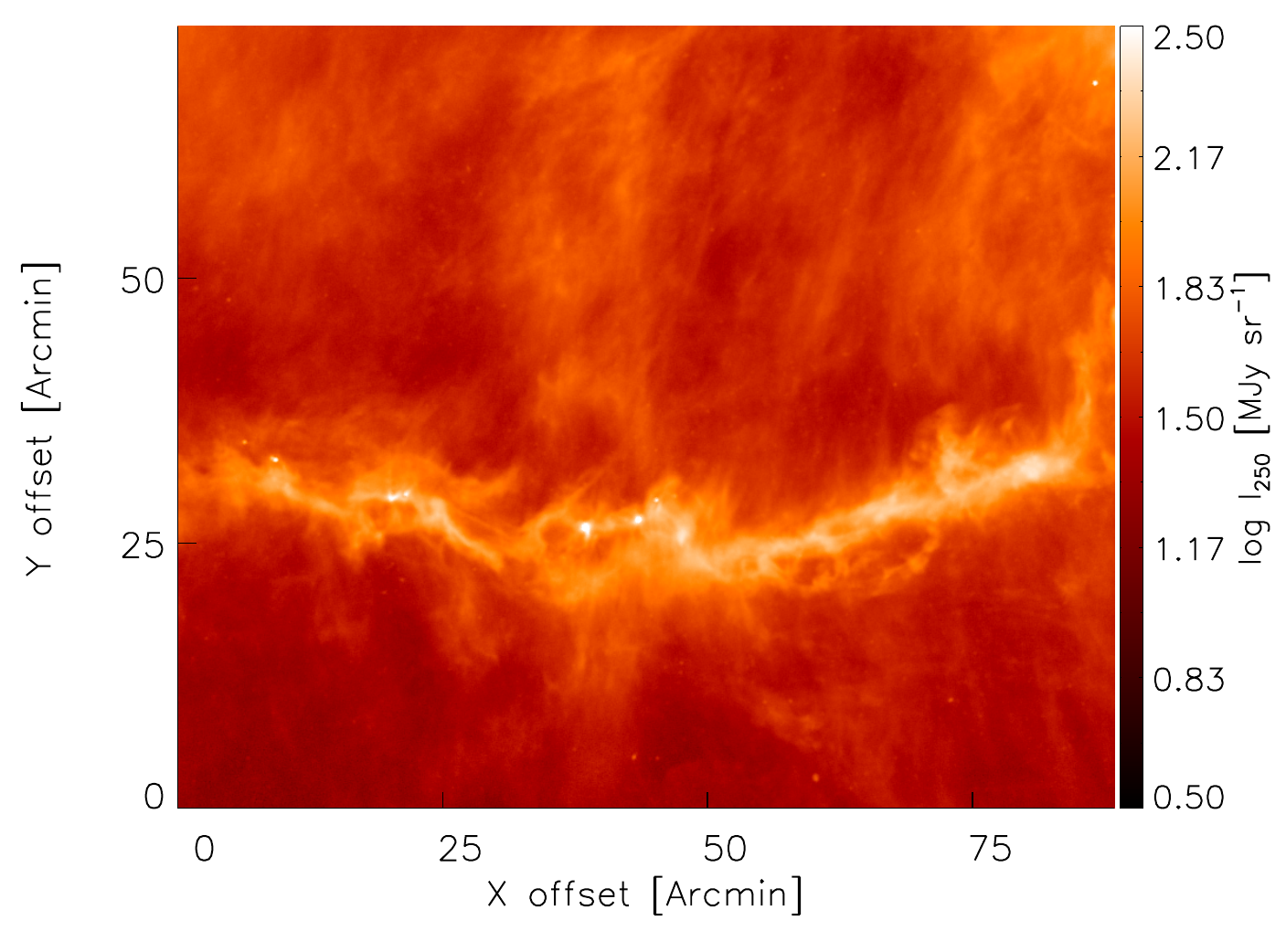}}%{Pipe_fig.ps}}
\hspace{0.3 cm }
\resizebox{0.49\hsize}{!}{\includegraphics[angle=0]{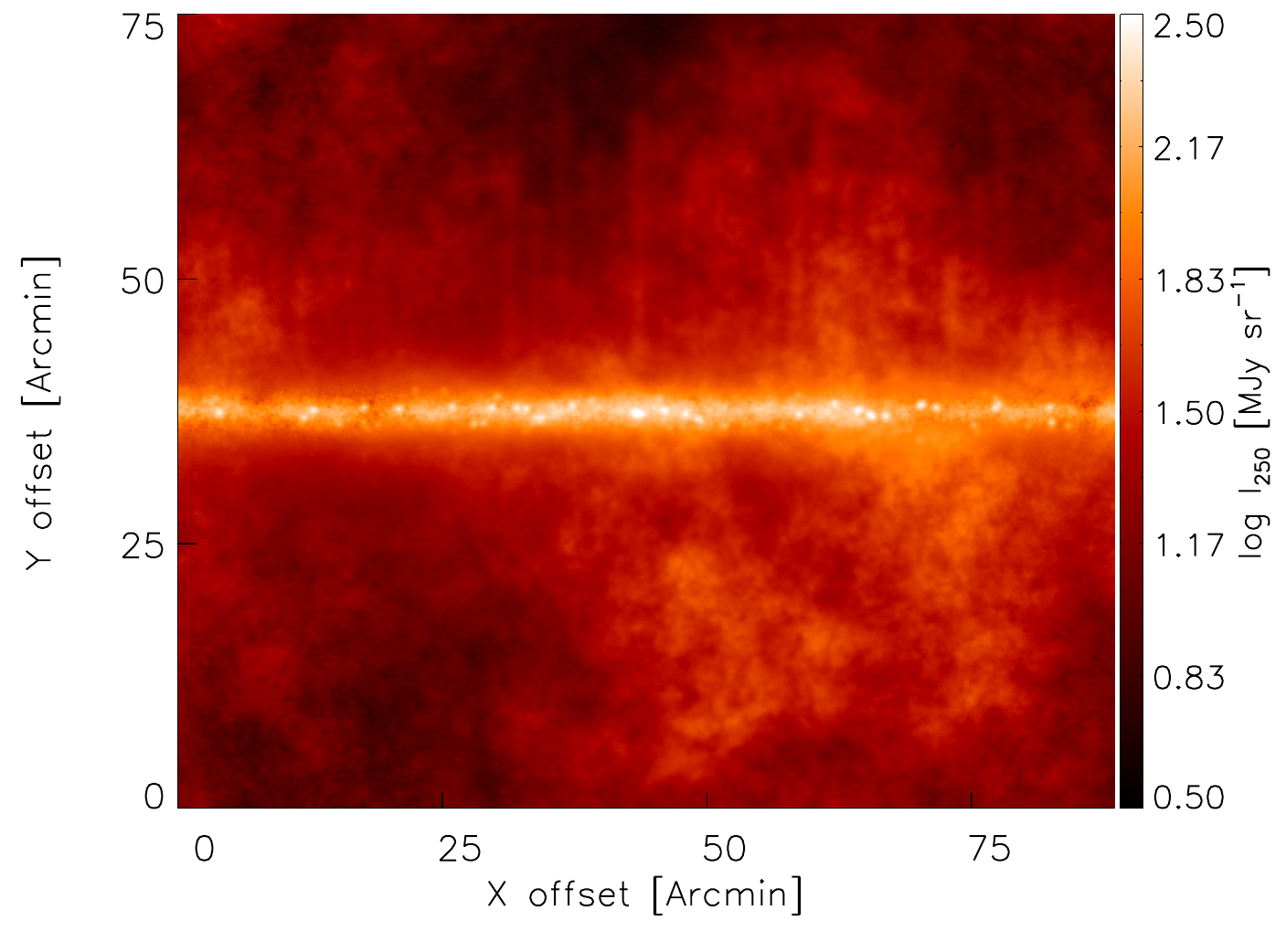}}
\caption{ {\bf Left:} {\it Herschel}/SPIRE 250-\micron\  image of the B211/B213 region in Taurus at the native beam resolution of 18.2\arcsec, but rotated in equatorial coordinates in clockwise direction by 37.4$^\circ$.
(see \citealp{palmeirim2013}).
%The image was gridded at a pixel size of 3.2\arcsec. 
{\bf Right:} Fully synthetic image mimicking the main features of the real image shown in the left panel, and resulting from
the co-addition of a synthetic filament image and a synthetic background image.
The synthetic filament image was based on the Plummer-like model of the B211 filament reported by \cite{palmeirim2013}: flat inner
radius $R_{\rm flat} =0.035$ pc,  contrast $\delta_{\rm c} \sim 6$, and power-law index $p$=2 at large  radii.
The background image was modeled as non-Gaussian cirrus fluctuations with a logarithmic power spectrum slope of $-3$ (see text for details),  
plus low-contrast filamentary structures resembling striations. 
The synthetic striations were placed such that their long axis 
is perpendicular to the main filament at a regular separation of 0.1 pc.}
\label{fig-simB211}
\end{figure*}

It can be seen in  Fig.~\ref{Fig-Aquila-bkg}b that the amplitude of the power spectrum 
arising from the population of synthetic filaments (blue curve) is lower 
than the amplitude of the power spectrum of the Aquila original image (green
dashed curve) by a factor of $\sim$ 5 at $k\sim k_{\rm fil} \sim (0.6/\theta_{\rm  fil})$
$ \sim 0.45$ arcmin$^{-1}$, corresponding to the characteristic angular width of the synthetic filaments, 
$\theta_{\rm fil} = 79\arcsec $ (i.e., 0.1 pc at $d$~=~260 pc). 
Clearly, the power contribution of the 
%population of 
synthetic filaments is not
strong enough to be detected in the power spectrum.   
The residuals of the best power-law fit with respect to the
power spectrum of the Aquila original image are shown as black triangles in Fig.~\ref{Fig-Aquila-bkg}c
%Fig.~\ref{Fig-Aquila}c 
as a function of angular frequency.  The red solid circles in  Fig.~\ref{Fig-Aquila-bkg}c 
%Fig.~\ref{Fig-Aquila}c
represent similar residuals for the Aquila background plus synthetic filament image. 
Based on this simulation, we conclude that the injection of a population of synthetic 
filaments with a distribution of column density contrasts similar to that
observed in the real Aquila image does not have any significant effect
on the shape of the power spectrum.

%%\subsection{Effect of synthetic population of filaments on Aquila background image}
%\textbf{In a manner similar to the Polaris case described in Sect.~\ref{Sect:polaris-bkg}, 
%the plots  shown in Fig.~\ref{Fig-Aquila-bkg} summarize the effect of a population of synthetic 0.1~pc filaments added 
%to a filament-subtracted background image of the Aquila region. 
%Like in the Polaris case, the background image of the Aquila cloud was obtained from the {\it Herschel}/SPIRE $250\, \mu$m 
%image after removing observed filaments using the \textsl{getfilaments} algorithm \citep{Menshchikov2013}. 
%The subtraction of real filaments from the background image hardly enhanced the relative contribution of the synthetic filaments to the power spectrum.
%}

In Appendix \ref{Appen:extreme}, we also explore a more extreme case where 
the distribution of column density contrasts for the injected synthetic filaments is similar 
in shape to the distribution shown in Fig.~\ref{fig-contrast-dist}, but 
with higher mean contrast $<\delta^{\rm  peak}_{\rm c}>$ $=$ 2.7 and 
maximum contrast $\delta^{\rm  max}_{\rm c}$ =15. 
In this extreme case, the population of synthetic filaments is strong enough to produce 
a detectable signature in the resulting power spectrum.

%\begin{figure}
%\centering
% \resizebox{\hsize}{!}{\includegraphics[angle=0]{./Plots/AquilaImgPowspecBeamNoiseCorrnfil_100_delta_mmp0.30_3.00_1.00DAF_5.97.ps}}%./Plots/AreaFilling_Contrast_20x20.ps}}%{Pipe_fig.ps}}
% \caption{$\chi^$-variance of residuals as a function of 
%$\chi^2$-variance of residuals as a function of
%   column density contrast ($\delta_c$) and
%   area filling factor ($f_{\rm fil}$).
%   Cross and plus symbols show the position of
%   Aquila and Polaris molecular clouds in the $\delta_c - f_{fil}$ plane.
%   At low filament contrast and
%   low area filling fraction, the synthetic filaments produce an
%   undetectable deviation (hence, low variance) in power spectrum whereas for high contrast
%   and high area filling fraction the deviation of power spectrum near
%   the angular scale of. }%$k_{\rm fil}$ is significant.}
%\label{fig:deltaAfil}
%\end{figure}

%\section{Image with a single, prominent filament such as B211 in  Taurus} \label{appen:taurus}
\section{Power spectrum of synthetic data with a single, prominent filament} \label{sect:taurus}

 We also examined the power spectrum of  an image 
%a {\it Herschel} image 
with a single dominant filament such as the \her /SPIRE 250~\micron\ image of the
B211/B213 region in the Taurus cloud at $d \sim 140\,$pc (Fig.~\ref{fig-simB211}a)\footnote{In order to capture the largest possible rectangular area, 
we rotated the SPIRE map by $37.4^\circ $ in the clockwise direction with respect to an equatorial frame.}.
%observed by \her\  as part of HGBS survey \citep{palmeirim2013}.
For the present purpose, we only used a 1.2$^{\circ}$ $\times $1.0$^{\circ}$ 
portion of the original SPIRE image of B211/B213, where a single filament dominates over a
length scale of $> 1.5^{\circ}$ (or $>4\, $pc).
\citet{palmeirim2013} studied the column density structure of the B211/B213 filament in detail 
and found that it is accurately described by a Plummer-like cylindrical density distribution 
with flat inner radius $R_{\rm flat} \sim 0.035\, $pc and power-law index $p = 2\pm0.2$ at larger radii up to an outer radius $R_{\rm out}\sim 0.4\, $pc.
Moreover, \citet{palmeirim2013} suggested that the Taurus main filament accretes mass from the ambient cloud 
through a network of lower-density 
%filamentary structures, called 
striations, observed roughly perpendicular to the main filament.
%nearby diffuse gas reservoir through low-column density filaments called  striations. 
%These striations are perpendicular to the B211 filament and parallely  aligned along the magnetic field.
%
Based on these findings, 
%We adopted these parameters to 
%Adopting these parameters, 
we constructed a synthetic image of a Plummer-like filament of length $\sim 4 \, $pc, 
with the same Plummer parameters as quoted above, and positioned horizontally 
in a $\sim$ 1.5$^{\circ}$ $\times $1.5$^{\circ}$ two-dimensional box. 
The  
%column density 
contrast of the synthetic filament was chosen to be $\delta_c$ $\sim$ 6, a value close to the observed contrast 
of the B211/B213 filament in the SPIRE $250\ \mu$m image
%(and only slightly lower than the column density contrast $\sim $ 10 measured for the B211 filament -- see Table B.1 of 
(see \citealp{palmeirim2013}). 
To mimic the observations, we added a population of synthetic cores with Bonnor-Ebert-like radial profiles randomly distributed along the filament. The flat inner radius R$_{\rm flat}$ of the cores was fixed to a constant value of 0.02 pc.
In order to create a synthetic background image similar to the real data, 
%to that of Taurus  
we carefully studied the statistical properties of the \her\  250~\micron\ image in the vicinity of the Taurus main filament. 
We selected a rectangular field
%coverage area
to the north of the B211/B213 main filament such that the nearest edge of the field was at least 0.2~pc away from
the filament crest. 
%of the main filament.
We then evaluated the power spectrum of this field and found 
%Analysis of the power spectrum analysis revealed 
a logarithmic slope  $\gamma \sim -3.0 \pm 0.2$.
A purely synthetic background image was next  
generated using a non-Gaussian fractional Brownian
motion (fBm) technique (\citealp{mamd2003}) with positive values and statistics such that the power spectrum
of the background field had a logarithmic slope similar to that of the Taurus background field ($\gamma_{\rm back} =- 3.0$).
To make the synthetic background image more similar to the Taurus observations,
%match our simulated background image with the observations 
we also inserted a distribution of lognormally distributed low-contrast  (0.1 < $\delta_c$ < 0.5) filamentary structures with Gaussian profiles perpendicular 
to the main filament as a proxy for the observed striations. We placed perpendicular striations at a regular separation of $\sim$ 0.1 pc to match the observations of \cite{tritsis2018}.  
The width of these synthetic striations 
%in the simulation 
was fixed to 0.08~pc.

\begin{figure}
  \resizebox{\hsize}{!}{\includegraphics[angle=0]{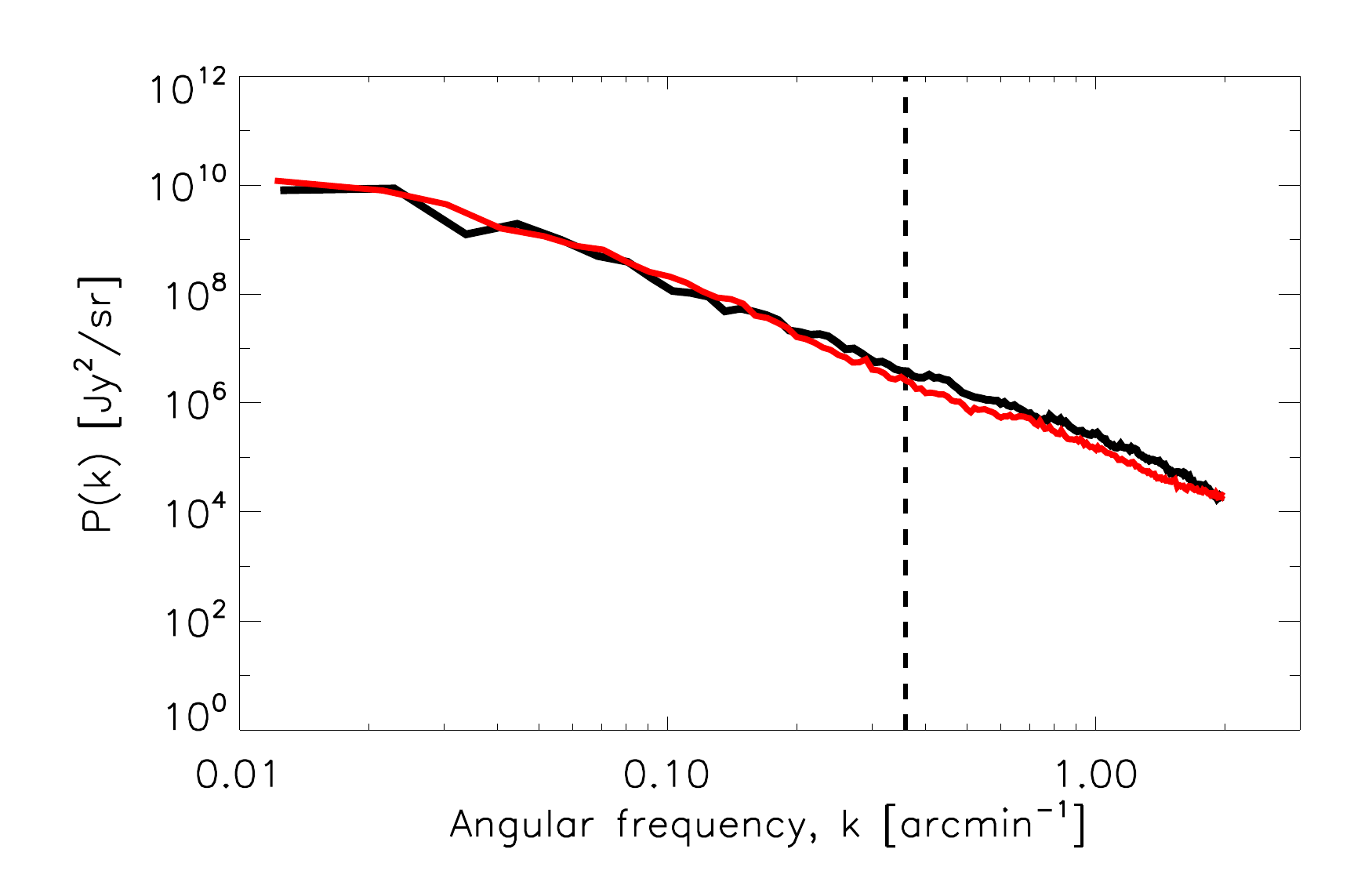}}%{Pipe_fig.ps}}
 \caption{Comparison of the power spectrum of the {\it Herschel}/SPIRE 250-\micron\ image of Fig.~\ref{fig-simB211}a 
(black solid curve) with that of the synthetic image of Fig.~\ref{fig-simB211}b (red curve).
Note the absence of any significant feature around $k \sim k_{\rm fil}$ (vertical dashed line) in both power spectra.
}
\label{fig-Taurus-ps}
\end{figure}

The final 
%synthetic 
image, obtained after co-adding all three synthetic image components (background, striations, and main 
filament with embedded cores), is shown in Fig.~\ref{fig-simB211}b. 
For reference and comparison with the synthetic data discussed in Sect.~\ref{Sec:polaris} and Sect.~\ref{Sec:Aquila}, 
this image has $\delta_c^2 A_{\rm fil} \sim 0.125$. 
Figure~\ref{fig-Taurus-ps} compares the power spectrum of the synthetic image (red curve) 
with that of the \her /SPIRE 250~\micron\ image (black solid curve).
The vertical dashed line in Fig.~\ref{fig-Taurus-ps} marks the angular
frequency $k_{\rm fil}$ corresponding to a linear scale of $\sim 0.1\,$pc, i.e., roughly the inner width of both the synthetic filament 
and the B211/B213 filament.
%a filament width of $\sim 0.1\,$pc.
Clearly, like for the other two regions considered in this paper, the power spectrum of the Taurus B211/B213 data 
does not reveal any 'kink' or 'break' 
%in the power spectrum 
at frequencies close to $k_{\rm fil}$.  Furthermore, this is also the case for the synthetic data of
Fig.~\ref{fig-simB211}b, despite the presence of a prominent cylindrical filament with $\sim 0.1\, $pc inner diameter.
%Does this indicate a scale-free nature of molecular
%filaments? We look into this problem more closely through simulations.
%
This further illustrates how the characteristic scale of embedded structures may be hidden
and undetectable in a global power spectrum.

	\section{Combined  effect of filament contrast and area filling factor} \label{sec:deltaAfil}

	%The magnitude/amplitude of excess power relative to the
        %best-fit power law model power spectrum at the characteristic
        %frequency ~$k_{\rm fil}$ depends upon the combined effect of
        %the mean filament contrast in the image and the fraction of area
        %covered by the filaments (area filling factor, $A_{\rm fil}$).
        %The area filling factor $A_{\rm fil}$ is defined as $
        %\Sigma_{i=1}^{N_{\rm fil}} L_{\rm i} \times W_{\rm fil}$/$A$,
        %where $L_{i}$ is the length of the $i^{\rm th}$ filament, and
        %$W_{\rm fil}$ ($\sim$ 0.1 pc) is the lateral width of
        %filaments and $A$ is the total coverage area in the image
        %being analyzed, and $N_{\rm fil} $ is the total number of
        %synthetic filaments injected in the image.  In order to
        %quantify the magnitude of the deviation from the best power-law
        %fit we have used $\chi^2$-variance as our metric.  We
        %calculated the variance of the residuals in the vicinity 
        %of $k_{\rm fil}$ where the
        %contribution of filament power is expected to be maximum. The
        %variance is defined as 
        %\begin{equation}
        %\chi^2_{\rm variance} = \Sigma_{k_{\rm min}} ^{1.5 k_{\rm
        %    fil}}Res(k)^2/N_{\rm freq}, 
       % \end{equation} 
        %where $Res(k)$ is the
        %residual at angular frequency $k$, and $N_{\rm freq}$ is the total number
        %of frequency modes between $k_{\rm min}$ and $1.5\times k_{\rm
        %  fil} $. For low $A_{\rm fil}$ and $\delta_{c}$ the variance
        %is very small while for high $A_{\rm fil}$ and high
        %$\delta_{c}$ this metric is relatively very high. 

        In order to further explore the dependence of the total power spectrum on filament contrast  and area filling factor,         
        we performed two separate grids of 20$\times$20 Monte-Carlo simulations based on two different sets of synthetic filament populations, 
        one with Gaussian radial profiles and the other with Plummer-like profiles with $p$=2 (see Appendix~\ref{Appen:plummer}). 
        The simulated images  spanned a broad range of average filament
        %density 
        contrasts $<\delta_c>$ and filling factors $A_{\rm fil}$.   
        In practice, we injected a fixed number of synthetic filaments of 0.1~pc width in the \her /SPIRE 250\ \micron\ image of Polaris and 
        controlled the area filling factor by varying the length of the filaments. 
        %Since the amplitude of the power spectrum is directly proportional to
       %the square of over-densities, it is expected intuitively that the excess 
      %$\chi^2_{\rm variance }$ in the residuals  should be correlated with $\delta_c^2$ 
     %of the synthetic filaments, and should linearly proportional to the $A_{\rm fil}$. 
     For each realization, we then calculated the $\chi^2_{\rm variance}$ of the residuals between the best power-law fit and the 
     net output power spectrum,  as described in Sect.~\ref{sec:diagnostic}. 
     
  \begin{figure}[!hbp]
  \resizebox{1.0\hsize}{!}{\includegraphics[angle=0]{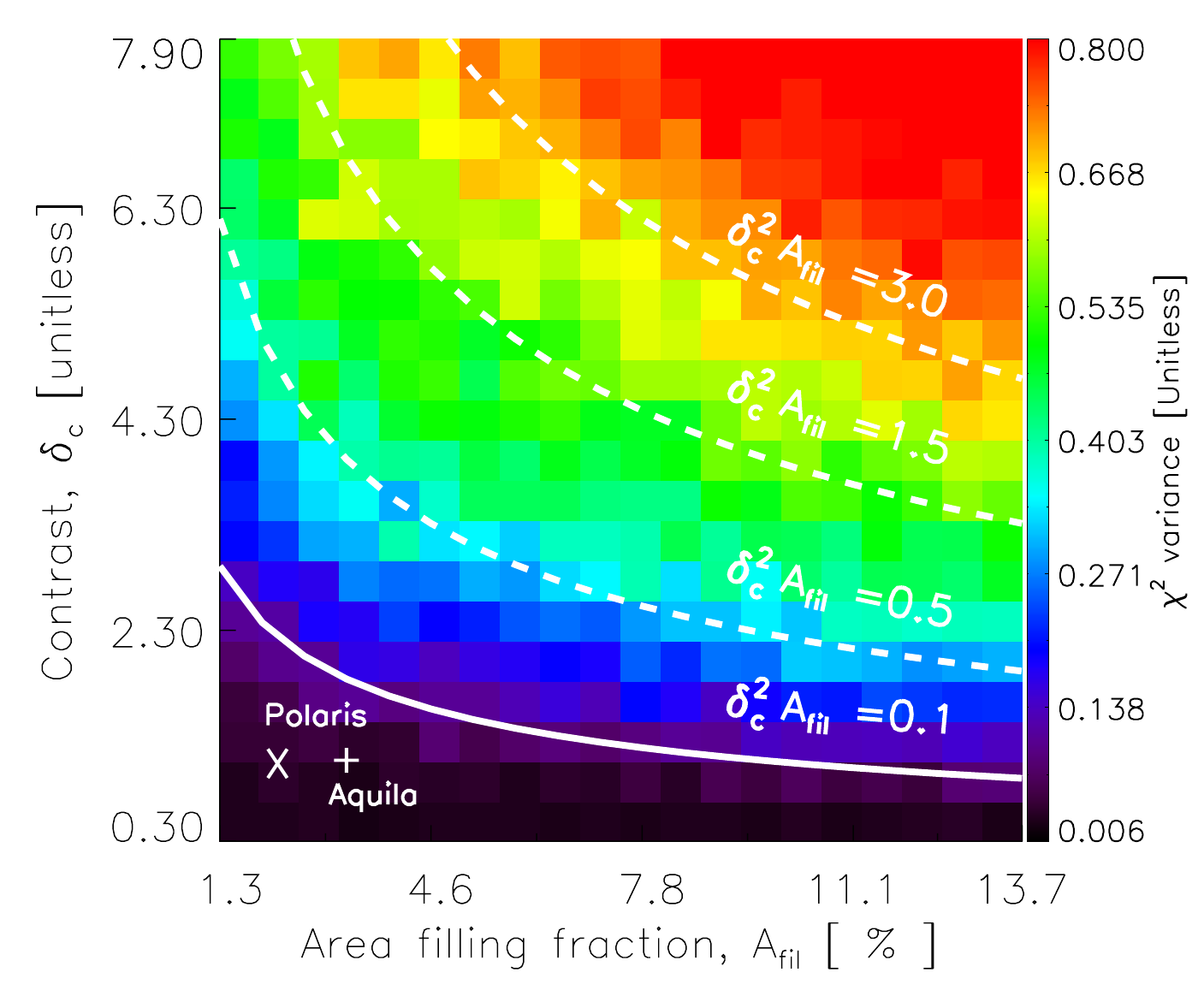}}%{Pipe_fig.ps}}
 \caption{Map of the $\chi^2$-variance of the residuals between the best power-law fit  and the output power spectrum 
 as a function of  column density contrast ($\delta_c$) and area filling factor ($A_{\rm fil}$) in a grid of  
 simulations based on a set of $20\times 20$ populations of Gaussian synthetic filaments, all 0.1~pc in width 
 (see text of Sect.~\ref{sec:deltaAfil} for details).
The white solid curve marks the fiducial limit $\delta_c^2A_{\rm fil } =0.1$ above which 
the effect of the characteristic filament width can be detected in the power spectrum 
(see Fig.~\ref{fig:deltaAfil} below). 
%The dashed contours show higher levels of $\delta_c^2A_{\rm fil } $= 0.3 and 0.6. 
 % 
The white plus and cross symbols mark the positions of the observed populations of filaments 
in the Aquila and Polaris clouds, respectively \citep[cf.][]{arzoumanian2017}.
}
\label{fig:delta2fil}
\end{figure}

\begin{figure}[t]
  \resizebox{\hsize}{!}{\includegraphics[angle=0]{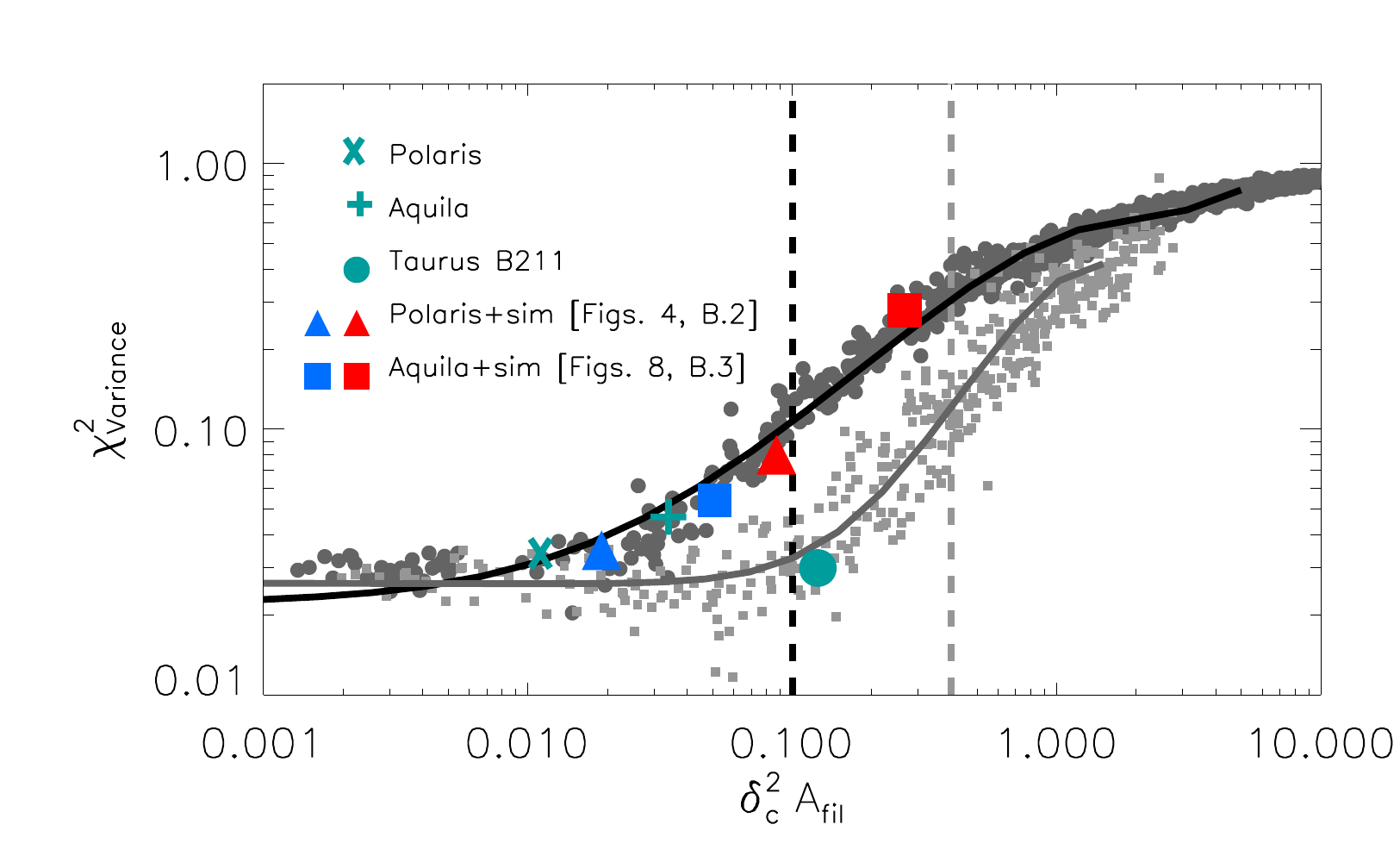}}%{Pipe_fig.ps}}
 \caption{$\chi^2_{\rm Variance}$ of the residuals between the best power-law fit  and the output power spectrum
 as a function of $\delta_c^2A_{\rm fil}$.  The black solid circles represent the same set of simulations with Gaussian filaments as in Fig.~\ref{fig:delta2fil}. 
% The green solid circles are the variance measure 
%   for a set of simulations with a varying column density contrasts $\delta_c$ and  area filling factors $A_{\rm fil}$.
The light gray squares represent our set of simulations with Plummer-like ($p=2$) filament profiles. 
The corresponding black and gray curves are polynomial fits to guide the eye.
  % 
% 
 % At low $\delta_c^2A_{\rm fil}$ the $\chi^2$-variance is constant because the contribution of synthetic filaments to
%the power spectrum is very low. 
The black and gray vertical dashed lines mark the fiducial limits of  $\delta_c^2A_{\rm fil}\sim0.1$ and $\delta_c^2A_{\rm fil}\sim 0.4$ 
above which Gaussian-shaped and Plummer-shaped filaments become detectable in the residual power spectrum plot, respectively.
%Above these fiducial limits, Gaussian and Plummer-like filaments become detectable in the residual power spectrum plot 
%(for Gaussian filaments see Figs.~\ref{fig:polaris-0.5}c, \ref{fig-Polaris-extreme}c, and \ref{fig-Aquila-extreme}c).
The green cross, plus, and solid circle symbols mark the positions of the Polaris, Aquila, and Taurus clouds, respectively,  
based on the comprehensive study of filament properties by \citet{arzoumanian2017}. 
%The green solid circle shows the position of Taurus (larger than the Gaussian fiducial threshold, but lower than the Plummer fiducial detection limit) .
%
%the $\chi^2$-variance is almost 3$\times$ $\chi^2$-variance of the background image. 
}
   \label{fig:deltaAfil}
\end{figure}

     Figure~\ref{fig:delta2fil} summarizes the dependence of $\chi^2_{\rm variance}$ on $<\delta_c>^2 $ and $A_{\rm fil}$  for Gaussian synthetic filaments.
     The map of the $\chi^2_{\rm variance}$ as a function of  $<\delta_c>^2 $ and $A_{\rm fil}$ is qualitatively similar for Plummer-like synthetic filaments.
     Figure~\ref{fig:deltaAfil} 
     %demonstrates 
     shows that there is a  tight correlation between $\chi^2_{\rm variance}$ and $<\delta_c>^2 \times \, A_{\rm fil}$ 
     for both  Gaussian (black solid circles) and Plummer-like filaments (gray solid squares), as expected from Eq.~\ref{eq:phi}. 
In both cases,  $\chi^2_{\rm variance}$ appears to be a non-linear function of $\delta_c^2 A_{\rm fil}$, with a flat 
portion at low $\delta_c^2 A_{\rm fil} $ values (i.e., $\delta_c^2 A_{\rm fil} \la 0.02$ for Gaussian filaments, 
$\delta_c^2 A_{\rm fil} \la 0.07$ for  Plummer-like filaments), a rising portion at higher $\delta_c^2 A_{\rm fil} $ values, 
with an inflection point close to $\delta_c^2 A_{\rm fil} \sim  0.1$ in the Gaussian case
and $\delta_c^2 A_{\rm fil} \sim  0.4$ in the Plummer case (see Fig.~\ref{fig:deltaAfil}). 
It can also be seen that, for the same value of $\delta_c^2 A_{\rm fil}$, the $\chi^2_{\rm variance}$ is  lower 
for Plummer synthetic filaments than for Gaussian synthetic filaments\footnote{A Plummer-like filament with $p\la 2.6$ 
contributes less power to the power spectrum at low angular frequency than a Gaussian filament with similar inner width
and contrast (see Fig.~\ref{fig-kernel-ps}b). Therefore, at $k<k_{\rm fil} $, Plummer-like filaments with $p\la 2.6$ lead 
to a lower overall $\chi^2_{\rm Variance}$ compared to Gaussian filaments.}.
             
Qualitatively, this behavior may be understood as follows. 
At low $\delta_c^2 A_{\rm fil} \la 0.02$ values (or $\delta_c^2 A_{\rm fil} \la 0.07$  for Plummer-like filaments), 
the contribution of synthetic filaments to the total power spectrum is negligible, and 
$\chi^2_{\rm Variance}$ is dominated solely by the residuals of the original background image. 
Therefore,  $\chi^2_{\rm variance}$ retains the value $\chi^2_{\rm variance,bkg}$ it has for the original image 
and remains constant despite the addition of synthetic filaments. 
As $\delta_c^2 A_{\rm fil}$  increases, the $\chi^2_{\rm variance}$ for both Gaussian and Plummer filaments also increases, 
reaching a value of about $3\times\chi^2_{\rm variance, bkg} \sim 0.1$ at $\delta_c^2 A_{\rm fil} \sim 0.1 $ (Gaussian case) 
or 0.4 (Plummer case). 
%
%$\delta_c^2 A_{\rm fil} > 0.02$,  $\chi^2_{\rm variance}$ also increases, but, 
%its effect is not entirely prominent until $\delta_c^2 A_{\rm fil} \sim 0.1 $. 
%At $\delta_c^2 A_{\rm fil} = 0.1 $, the value of $\chi^2_{\rm variance}$ 
%is about $3\times\chi^2_{\rm variance, bkg}$ $\sim$ 0.1.  
We take these values of $\delta_c^2 A_{\rm fil}$ as fiducial limits for the detection of a characteristic filament width 
in the image power spectrum for Gaussian- and Plummer-shaped filaments, respectively. 
These fiducial detection limits are marked by black and gray vertical dashed lines in Fig.~\ref{fig:deltaAfil}. 
%for Gaussian and Plummer-shaped  ($p=2$) filaments, respectively. 

%\begin{figure}[t]
%  \resizebox{\hsize}{!}{\includegraphics[angle=0]{./Plots/fil_index.ps}}%{Pipe_fig.ps}}
% \caption{ Correlation coefficent averaged over piece-wise sectors of $\chi^2$ versus $f_{\rm fil}^{\alpha} \delta_c^2$ scatter
%plot. For $\alpha =0.8$, the curve has maximum correlation.}
%   \label{fig:index-correl}
%\end{figure}

%To put this $\chi^2_{\rm variance}$ relation in context and to give a perspective, 
To put the simulation results shown in Fig.~\ref{fig:delta2fil} and Fig.~\ref{fig:deltaAfil} in context, 
we recall that the Polaris simulation of
Fig.~\ref{fig:polaris_filremoved-0.5}a in Sect.~3 
%the realization of synthetic filaments for the case of Polaris shown in Fig.~\ref{fig:polaris-0.5}a 
had $\delta_c^2 A_{\rm fil}\sim$0.018 and  $\chi^2_{\rm variance}$ of 0.034 (see Fig.~\ref{fig:polaris-0.5}c), 
which is nearly the same as the observed $\chi^2_{\rm variance,bkg}$ $\sim$ 0.04 (see Fig.~\ref{Fig-polaris}). 
This particular simulation is marked by a blue triangle in 
the $\chi^2_{\rm variance}$-- $\delta_c^2 A_{\rm fil}$ plot of Fig.~\ref{fig:deltaAfil}.
%Similarly, 
The more extreme Polaris simulation presented in Fig.~\ref{fig-Polaris-extreme}a (Appendix~B),  for which there is a marginal detection of a characteristic scale 
in the residuals plot (see Fig.~\ref{fig-Polaris-extreme}c), has $\delta_c^2 A_{\rm fil} \sim 0.087$ and $\chi^2_{\rm variance} \sim 0.08$ 
(see red triangle in Fig.~\ref{fig:deltaAfil}).
%In Fig.~\ref{fig:deltaAfil}, this case is shown by the red triangle. 
Likewise, the blue and red square symbols in the $\chi^2_{\rm variance}$-- $\delta_c^2 A_{\rm fil}$ plot of Fig.~\ref{fig:deltaAfil} mark the positions  
of the two sets of Aquila simulations presented in Fig.~\ref{Fig-Aquila-bkg} and Fig.~\ref{fig-Aquila-extreme}, respectively.  
%In Fig.~\ref{fig:deltaAfil}, the blue and red square symbols, highlight the 
%positions in $\chi^2_{\rm variance}$ - $\delta_c^2 A_{\rm fil}$ plane for two sets of Aquila plus synthetic filament 
%realizations presented in Fig.~\ref{Fig-Aquila} and Fig.~\ref{fig-Aquila-extreme}, respectively.  
The red square in Fig.~\ref{fig:deltaAfil} has $\delta_c^2 A_{\rm fil}~\sim$ 0.27 (and $\chi_{\rm variance} \sim$ 0.28), significantly above the fiducial detection limit of 0.1, 
indicating that the signature of a characteristic filament width should be detectable in the power spectrum.
This is indeed confirmed by visual inspection of Fig.~\ref{fig-Aquila-extreme}b and Fig.~\ref{fig-Aquila-extreme}c.
 %  
%Figure~\ref{fig:delta2fil} shows the dependence  of $\chi^2_{\rm variance}$ parameterized as a function of filament contrast $\delta_c$ and 
%area filling factor $A_{\rm fil}$ based on the grid of simulations. The threshold for the detection of significant peaks for the combination of
%column density contrast $\delta_c$ and area filling factor $A_{\rm fil}$ is highlighted  by the solid 
%contour $\delta_c^2 A_{\rm fil}=0.1$. The successive levels of $\delta_c^2 A_{\rm fil}$= 0.3 and 0.6 are shown
%by the dashed contours. The cross symbol shows the position of Aquila based on the census of filaments and their properties. The weighed-average of column density contrasts and area filling factor of filaments in Aquila are $<\delta_c>$$\sim$ 1 and $A_{\rm fil}\sim$ %, respectively.  For Polaris,
%the average column density contrast, $<\delta_c>$, and area filling factor, $A_{\rm fil}$ are 1.27 and 0.4\%. 
Most importantly, for both Polaris and Aquila, the real  \her\ data lie in a portion of the $\chi^2_{\rm variance}$-- $\delta_c^2 A_{\rm fil}$ diagram 
where the filament contribution has a negligible impact on the power spectrum (see cross and plus symbols in Fig.~\ref{fig:delta2fil} and Fig.~\ref{fig:deltaAfil}). 
Also shown as a green filled circle in Fig.~\ref{fig:deltaAfil} is the locus of the \her\ data for the prominent filament system B211/B213 in Taurus (see Sect.~\ref{sect:taurus}), 
which has a very well characterized Plummer-like density profile with a power-law wing index $p=2\pm0.2$ \citep{palmeirim2013}.
It can be seen that the position of  the Taurus B211/B213 data in Fig.~\ref{fig:deltaAfil} is in excellent agreement with our set of simulations for 
Plummer-shaped filaments with $p=2$. 
Although the $\delta_c^2 A_{\rm fil} \sim 0.125$ value of the Taurus B211/B213 data is greater than the fiducial threshold for Gaussian filaments,  
it remains much lower than the fiducial detection limit for Plummer ($p=2$) filaments.

 %It is an outlier in this plot because B211/B213 filaments are best described by  Plummer-like filaments (see Appendix ~\ref{appen:taurus}). 
 %In order to asses the effect of number of filaments on the $\chi^2$-variance plot we performed
%independent two sets of simulation with 50 filaments and 80 filaments, their results are shown
%in solid green circles and blue squares in Figure ~\ref{fig:deltaAfil}. 
We conclude that the essentially scale-free power spectrum of the {\it Herschel} images
observed toward molecular clouds such as Polaris, Aquila, or Taurus does not invalidate the existence of a characteristic filament width.
       
       % Figure~\ref{fig:deltaAfil} shows the metric of deviation
       % parametrized as a function of filament contrast ($\delta_{c}$)
       % and area filling factor $f_{\rm fil}$.  For $f_{\rm fil}\sim
       % 2\%$ and $\delta_c \sim$ 0.1 the variance was $\sim$ 0.005
       % whereas, on increasing $f_{\rm fil}\sim 20\%$ and $\delta_c
       % \sim$ 1.7, the magnitude of the variance increased by two
       % orders of magnitude.% $\sim$ 0.5.

%	\section{Concluding remarks}
	\section{Summary and conclusions}\label{Sec:conclusion}

We used numerical experiments to investigate the conditions under which the presence of a characteristic filament 
width can manifest itself in the power spectrum of cloud images. Our main findings and conclusions may be 
summarized as follows: 

%	Our numerical simulations show that the signature of
 %       characteristic inner width of the ISM filaments in power spectra
%        remains hidden. Its revelation, however, depends on various
 %       factors such as average contrast level of filaments with
 %       respect to the background, area filling factor, mean surface
  %      brightness of the sky and also on the distance of the target
 %       field.

	\begin{enumerate}

	\item The detectability of a characteristic filament scale in the power spectrum of an ISM dust continuum image primarily 
	depends on the parameter $\delta_c^2\, A_{\rm fil}$, where $\delta_c $ is the weighted average column density contrast 
	of the filamentary structures and $A_{\rm fil}$ their area filling factor in the image. A value $\delta_c^2\, A_{\rm fil}\gtrapprox 0.1$ 
	is required for the presence of a characteristic filament width to produce a significant signature in the power spectrum. 
	
	\item The \her\ Gould Belt survey images of nearby clouds typically have $\delta_c^2\, A_{\rm fil}\ll 0.1$ and therefore lie in a region 
	of the parameter space where filaments have a negligible impact on the power spectrum. Therefore, despite recent claims, the scale-free 
	nature of the observed power spectra remains consistent with the presence of a characteristic filament width $\sim 0.1\,$pc.
%        Most of the Gould-belt clouds have $\delta_c^2A_{\rm fil}\ll 0.1$.  Therefore, relating  
%        a continuous power-law of the ISM power spectrum with  a scale-free  physical process may not be a correct interpretation.  

         \item When the average filament contrast is low and/or when the filaments occupy a small area filling factor, the power spectrum 
         is dominated by the fluctuations of the diffuse, non-filamentary  component of the ISM. 
         
         \item Although a few filaments in the Polaris cloud have 
%         relatively high 
          column density contrasts up to $\delta_c \sim 0.9$, their area filling factor is extremely low $A_{\rm fil}\sim 2\%$, resulting 
          in a combined parameter $\delta_c^2\, A_{\rm fil} \sim 0.01$ for Polaris.
%          The parameter $\delta_c^2A_{\rm fil}$ for Polaris is $\sim$ 0.007, therefore, 
	The overall power spectrum of the \her\ images of Polaris is scale-free because the filaments are not contributing enough power 
	to produce a significant signature at the spatial frequency corresponding to the characteristic filament width of $\sim 0.1\,$pc.
          
      \item Despite the presence of several supercritical filaments of $\sim 0.1\,$pc inner width in the Aquila cloud, 
      the power spectrum of the Aquila \her\ images is also essentially scale free. 
      Due to the larger distance of the Aquila cloud compared to Polaris,  $\sim 0.1\,$pc filaments in Aquila
      subtend a smaller angular width scale on the sky, and therefore have a relatively low area filling factor. 
      Overall, our simulations suggest that the observed population of Aquila filaments contributes only $\sim 1/5$ of the total amplitude 
      of the power spectrum of the {\it Herschel} 250 \micron\ image.
      
%      \item The imprint of a characteristic filament width $\theta_{\rm  fil}$ in the
%        power spectrum is not confined to a narrow range of angular
%        frequencies, but is rather spread out over a range of frequencies $k \la k_{\rm fil}$ where 
%        $k_{\rm fil} \sim (0.6/\theta_{\rm  fil})$.
        
      \item Supercritical filaments with Plummer-like  profiles and high column density contrasts 
      lead to relatively small departures from a power-law power spectrum
      because the high contrast of the flat inner plateau in the density profile is compensated by broad power-law wings at large radii.  
      The B211/B213 filament system in Taurus, for example, despite having a very high central column density contrast, remains largely undetected in 
      the image power spectrum because of its Plummer-like density profile with $p \approx 2$.
      
%      which in particular, makes  it even more difficult to detect their presence in a power spectrum. This is because a Plummer profile with $p\lesssim 2.6$ has lesser 
    
    %Although  Plummer-shaped filaments  mostly have high column density contrasts,  that does not always facilitate to the detection of their presence in a power spectrum. This is because the power of  
      
  %    The presence of Plummer-shaped filaments ($p \lesssim2.6$) is even more difficult to detect  compared to Gaussian because power is transfered to 
%       hosts a significant
%       number of supercritical filaments it also shows a power
%        spectrum with a single power-law slope. Our simulations
%        suggest that the amplitude of the power spectrum of a
%        population of filaments with a similar contrasts distribution
%        as observed in the real Aquila image is lower than the power
%        spectrum of Aquila image by a factor of $\sim$ 5. 
         %Interestingly, slightly higher
        %filament contrasts alone cannot compensate for the decrease in
        %area filling factor.
                 
       \item 
We conclude that the scale-free appearance of the power spectra of cloud images 
does not invalidate the finding, based on detailed {\it Herschel} studies of the column density profiles,  
that nearby molecular filaments have a common inner width $\sim 0.1\,$pc \citep{arzoumanian2011, arzoumanian2017}.

	\end{enumerate}
	
\begin{acknowledgements}
This work has received support from the European Research Council 
under the European Union's Seventh Framework Programme 
(ERC Advanced Grant Agreement no. 291294 --  `ORISTARS').
We also acknowledge financial support from the French national programs 
of CNRS/INSU on stellar and ISM physics (PNPS and PCMI). 
A.R, and N.S., 
acknowledge support by the French ANR and the German DFG through 
the project "GENESIS" (ANR-16-CE92-0035-01/DFG1591/2-1).
P.P. acknowledges support from the Funda\c{c}\~ao para a Ci\^encia e a  
Tecnologia of Portugal (FCT)
through national funds (UID/FIS/04434/2013), from FEDER through COMPETE2020
(POCI-01-0145-FEDER-007672), and from the fellowship  
SFRH/BPD/110176/2015 funded by FCT
(Portugal) and POPH/FSE (EC). 
We are grateful to our colleague Alexander Men'shchikov for assistance with the \textsl{getfilaments} algorithm. 
This research has made use of data from the {\it Herschel} Gould Belt survey (HGBS) project (http://gouldbelt-herschel.cea.fr). 
The HGBS is a Herschel Key Programme jointly carried out by SPIRE Specialist Astronomy Group 3 (SAG 3), 
scientists of several institutes in the PACS Consortium (CEA Saclay, INAF-IFSI Rome and INAF-Arcetri, 
KU Leuven, MPIA Heidelberg), and scientists of the Herschel Science Center (HSC).
\end{acknowledgements}

	\bibliographystyle{aa}
	\bibliography{ref}

\begin{appendix}
 \section{Construction of synthetic filaments with Plummer-like density profiles}\label{Appen:plummer}

%\begin{figure}
% \resizebox{\hsize}{!}{\includegraphics[angle=0]{./Plots/AquilaImgPowspecBeamNoiseCorrnfil_100_contrast_0.50_.ps}}% Aquila_and_filament_powspecnfil_100_contrast_0.50_.ps}}
% \caption{{\bf a)}: \her\ dust emission image of Aquila at SPIRE
%   250~\micron\ plus Gaussian synthetic filaments image of width
%   $w_{\rm fil} \sim$ 0.1 pc.  The injected synthetic filaments have a
%   contrast of 0.5 compared to the local background. {\bf b) } The
%   solid black curve shows the power spectrum of Aquila (left image).
%   The red curve is the best fit model with a slope of $\gamma =
%   -2.16\pm 0.12$.  The vertical dashed line marks the angular
%   frequency scale corresponding to filament width of 0.1 pc at a
%   distance of 260 pc is $k_{\rm fil}\sim $ 0.38 arcmin$^{-1}$
%   (angular scale of Aquila filaments $\theta_{\rm fil-width}$
%   =79\arcsec). }
%\label{fig:Aquila-fil}
%\end{figure}

%\begin{figure}[t]
%  \resizebox{\hsize}{!}{\includegraphics[angle=0]{./Plots/AquilaImgPowspecBeamNoiseCorrnfil_100_contrast_1.00_.ps}}%{Pipe_fig.ps}}
% \caption{ Same as Fig. ~\ref{fig:Aquila-fil} but here the population
%   of synthetic filaments have a contrast $\delta_c \sim$ 100\%. The
%   red curve (right) is the best fit model with a slope of $\gamma =
%   -2.26\pm 0.13$.}
%\label{fig:Aquila-fil2}
%\end{figure}

\begin{figure}[!hbp]
  \resizebox{1.0\hsize}{!}{\includegraphics[angle=0]{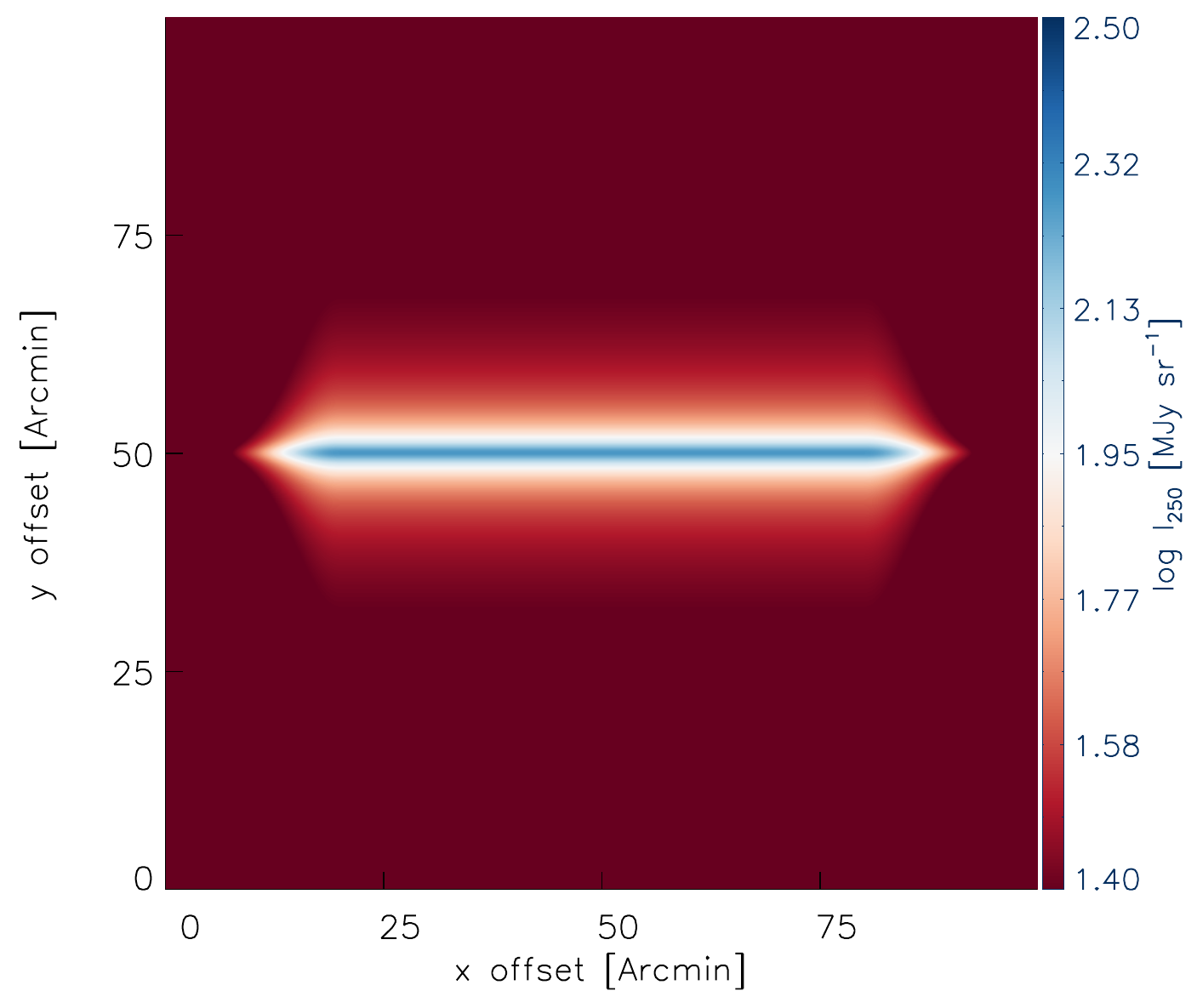}}%{Pipe_fig.ps}}
 \caption{Image of a synthetic filament with a Plummer-like transverse density profile 
    with a flat inner radius, $R_{\rm flat}=$ 0.03 pc, and a power-law wing with index $p=2$, 
  projected at a distance of 140~pc . In this example, the level of filament
   contrast was adjusted to $\delta_{\rm c} \sim$ 10.     
   }
\label{fig-plumm-fil}
\end{figure}

	%\section{Construction of synthetic filaments with a Plummer-like profile}
\noindent
        We adopted a slightly different technique to  produce filaments  with Plummer
	profiles  compared to the convolution technique used to generate  filaments with Gaussian profiles (see
	Sect.~\ref{sec:single-filament}). 
	%We numerically constructed 
	A Plummer-like  transverse profile was first constructed using the expression 
	%of a filament using the equation
	\begin{equation}
	K_{\rm Plummer} (r) = \frac{C}{\left[ 1+ \left( r/R_{\rm flat}
            \right)^2\right]^{(p-1)/2} },
	\end{equation}
	where $R_{\rm flat}$ is the flat inner width and $p$ is the logarithmic slope of the 
        power-law wing at large radii ($r >> R_{\rm flat}$). 
        %In the $p=2$ case, the
        %inner part of the plummer profile can be approximately fitted
        %with a Gaussian function of $FWHM = 3 R_{\rm flat}$.  
        In order to suppress the strong edge effect at the two ends of 
%        Plummer-like
        the model filament, we tapered both edges with a Gaussian function.
        Figure~\ref{fig-plumm-fil} shows an example of synthetic filament with a
        Plummer profile $p$ = 2 and $R_{\rm flat} = 0.1\,$pc, similar
        to the Taurus B211 filament  \citep{palmeirim2013}. 
        The power spectra of synthetic
        filaments with Plummer-like density profiles are discussed in Sect.~\ref{sec:single-filament}.

\section{Effect of extreme filament contrasts and area filling factors on the power spectrum} \label{Appen:extreme} 

 \begin{figure}
%  \centering
  \resizebox{ 0.925\hsize}{!}{\includegraphics[angle=0]{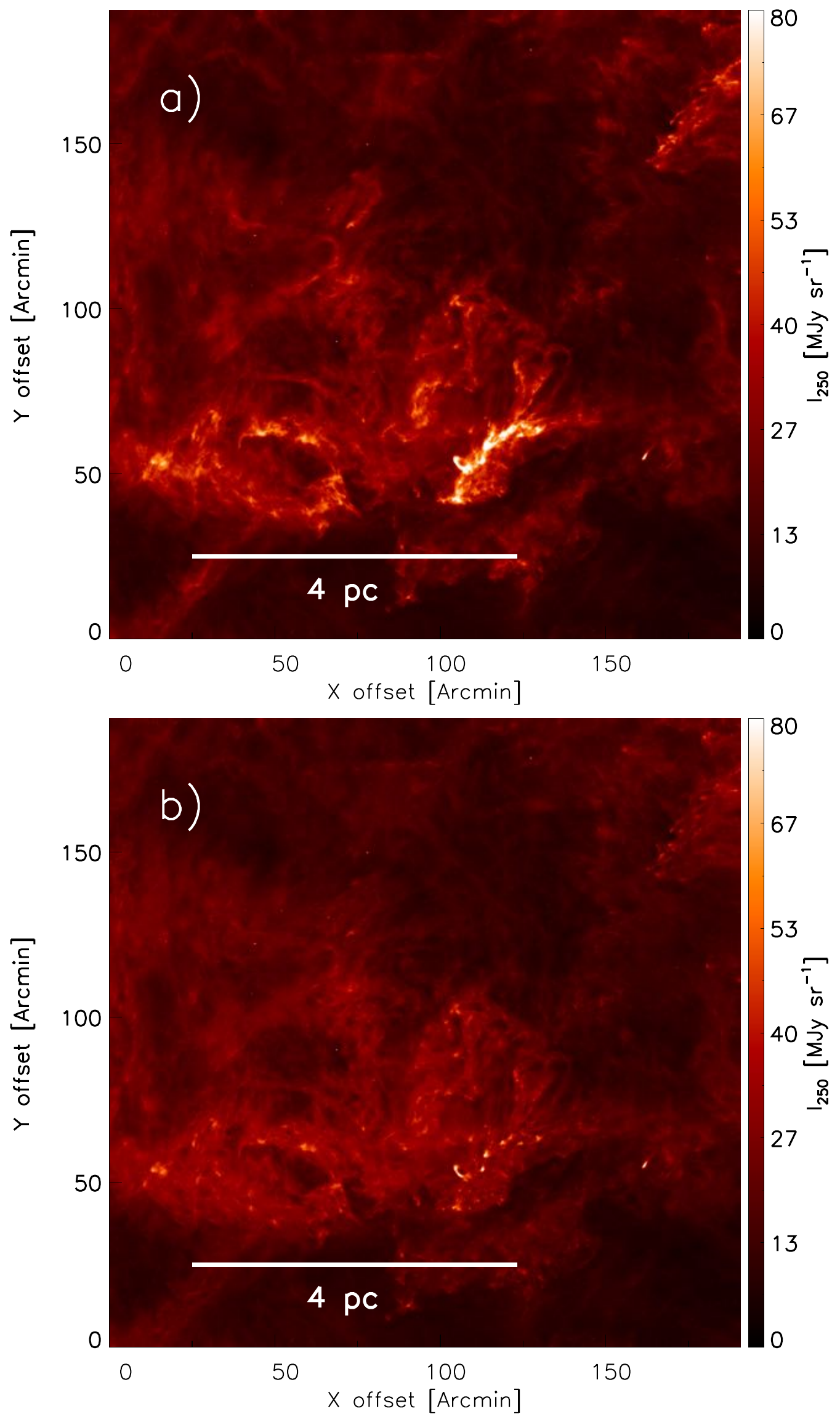}}%{Pipe_fig.ps}}

\caption{Comparison of  the  {\it Herschel} /SPIRE 250-\micron\  image of Polaris at the native beam resolution of 18.2\arcsec\ ({\bf a)}  --see \citealp{mamd2010}) 
with 
%The image was gridded at a pixel size of 3.2\arcsec. 
the filament-subtracted image of the same field ({\bf b)} panel) obtained with the \textsl{getfilaments} algorithm \citep{Menshchikov2013} and 
used as a ``filament-free'' background image in the numerical experiments shown in Fig.~\ref{fig:polaris_filremoved-0.5} and Fig.~\ref{fig-Polaris-extreme}. 
} 
\label{fig-simPolaris}
\end{figure}

\begin{figure}[!htp]
%  \centering
  \resizebox{0.95\hsize}{!}{\includegraphics[angle=0]{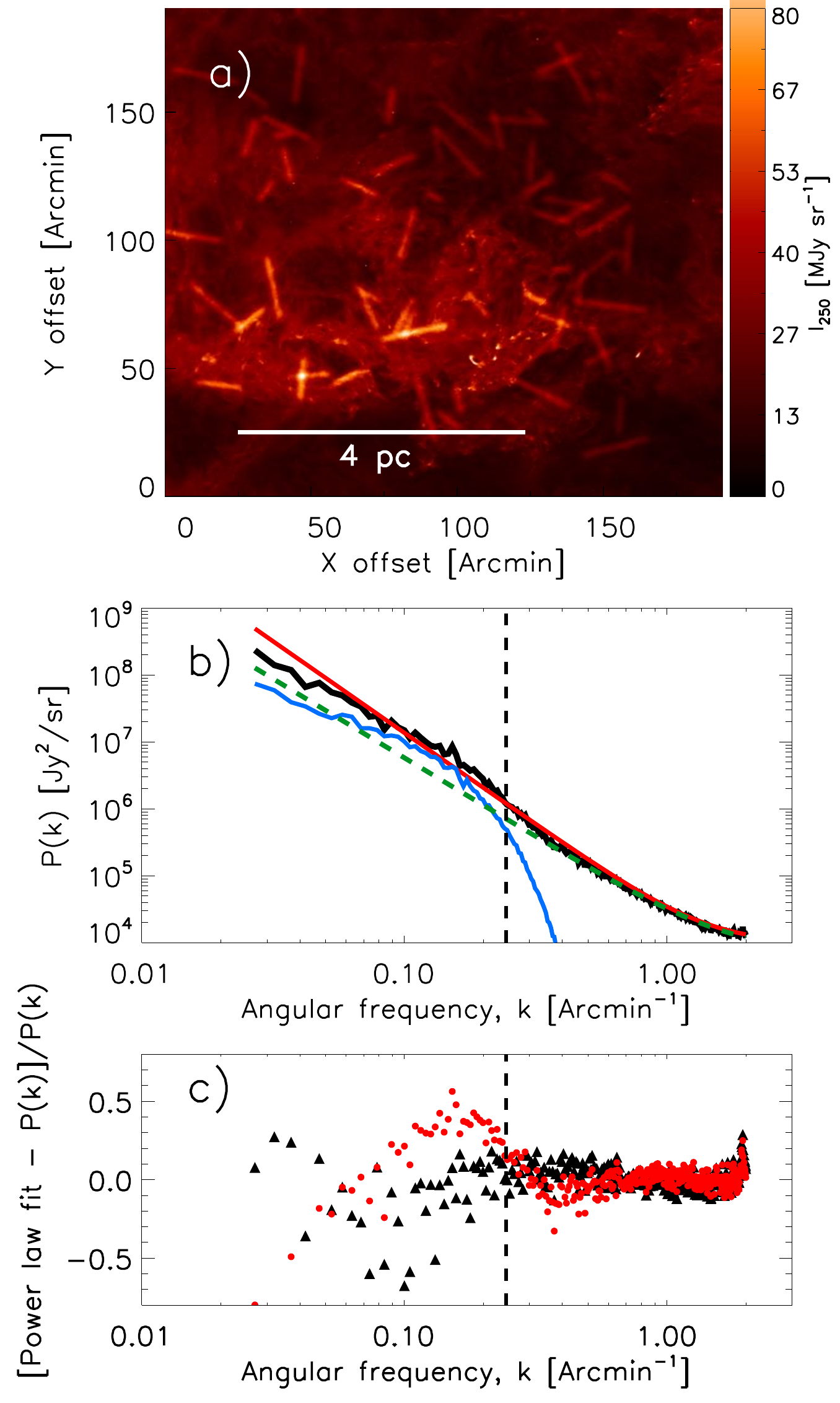}}%{PolarisImgPowspecBeamNoiseCorrnfil_70_contrast_1.10_DAF_7.97.ps}}%{Pipe_fig.ps}}
 \caption{Same as Fig.~\ref{fig:polaris_filremoved-0.5} but for a population of synthetic filaments 
with higher column density contrasts $\delta_c \sim 1.1$, resulting in $\delta_c^2A_{\rm fil} \sim 0.087$. 
(The area filling factor is similar to that in Fig.~\ref{fig:polaris-0.5}, $A_{\rm fil}\sim$ 7.2\%.) 
%{\bf a$)$} High contrast synthetic filaments clearly stands out compared to 
%background emission of Polaris and the parameter $\delta_c^2A_{\rm fil}$ is 0.085. 
In {\bf b$)$}, note how the amplitude of the power spectrum due to the synthetic filaments (blue curve) 
is comparable to that of the power spectrum of the Polaris original image (see green dashed line and Fig.~\ref{Fig-polaris}).
The logarithmic slope of the total power spectrum $P(k)_{\rm fil} +P(k)_{\rm Polaris}$ is $-$2.96. 
In  {\bf c$)$}, the plot of residuals between  the best power-law fit and $P(k)_{\rm Polaris + fil}$ (red solid circles) shows a peak near $k_{\rm fil}\sim$ 0.24 arcmin$^{-1}$. 
In this simulation, the $\chi^2_{\rm variance}$ is 0.08, close to the fiducial detection limit $\delta_c^2\, A_{\rm fil} \sim 1$  introduced in Sect.~\ref{sec:deltaAfil} (see Fig.~\ref{fig:deltaAfil}).     
   }
\label{fig-Polaris-extreme}
\end{figure}

\begin{figure}[!htp]
\resizebox{0.95\hsize}{!}{\includegraphics[angle=0]{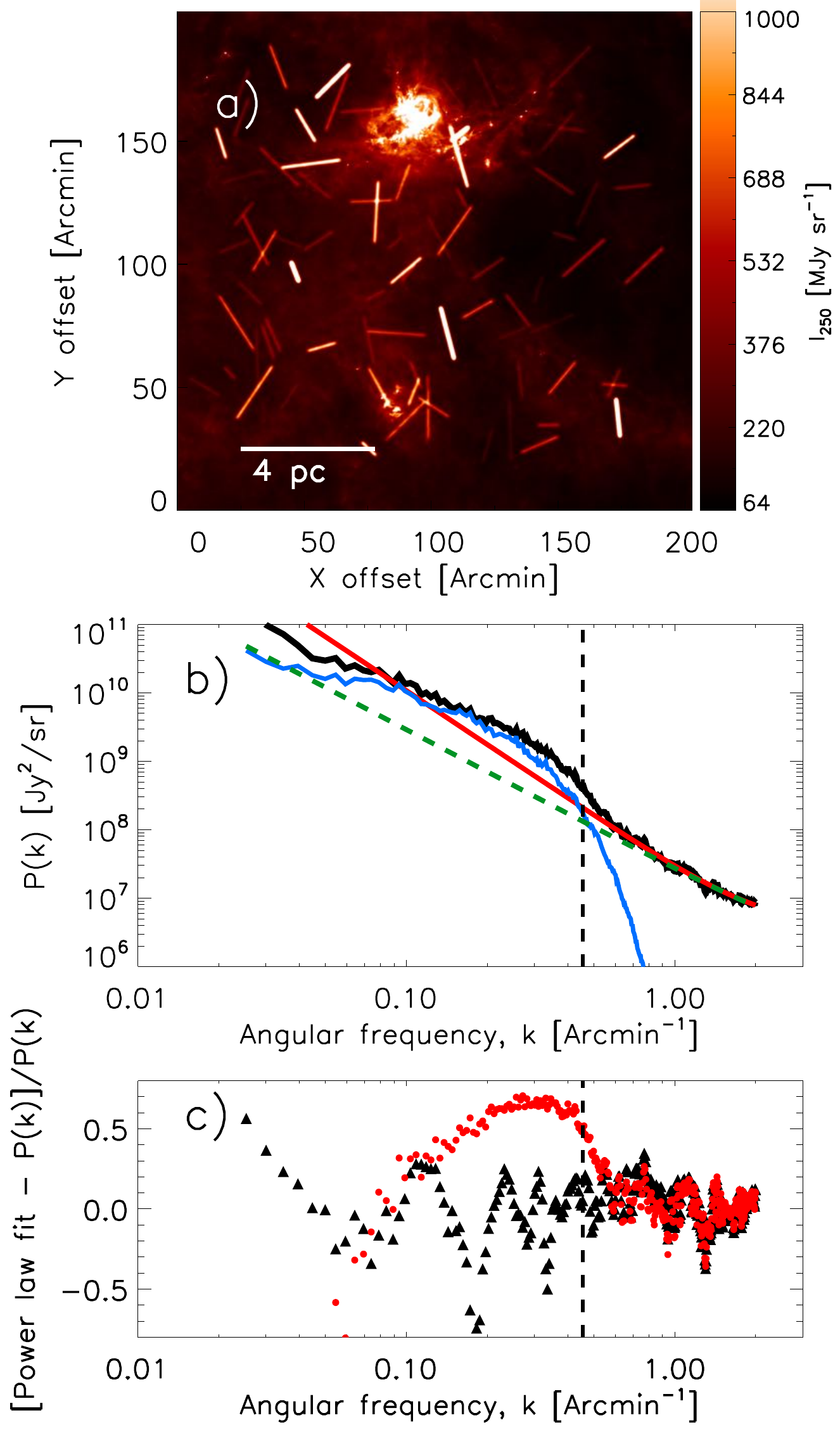}}% {AquilaImgPowspecBeamNoiseCorrnfil_100_delta_mmp0.40_15.0_2.30DAF_11.26.ps}}%{./AquilaImgPowspecBeamNoiseCorrnfil_100_delta_mmp0.30_15.0_1.50DAF_12.78.ps}} %AquilaImgPowspecBeamNoiseCorrnfil_100_delta_mmp0.30_15.0_1.50DAF_11.23.ps}}%AquilaImgPowspecBeamNoiseCorrnfil_100_delta_mmp0.30_10.0_1.50DAF_11.35.ps}}%{Pipe_fig.ps}}
 \caption{Same as 
{\bf Fig.~\ref{Fig-Aquila-bkg}}  
%Fig.~\ref{Fig-Aquila} 
but for a population of synthetic filaments with a more extreme (and unrealistic) distribution of 
column density contrasts corresponding to  $ <\delta_c> \sim 2.2 $ (see  Fig.~\ref{fig-contrast-dist-extreme}) and 
an area filling factor $A_{\rm fil} \sim 5.5\%$, resulting in a combined parameter $\delta_c^2\, A_{\rm fil} \sim 0.27$.  
In {\bf b$)$}, note how the power spectrum arising from the synthetic filament population (blue curve) 
dominates over the power spectrum of the Aquila original image (see green dashed curve and Fig.~\ref{fig:Aquila-img}). 
The logarithmic slope of the best power-law fit to the total power spectrum $P(k)_{\rm fil} +P(k)_{\rm Aquila}$ is $- 2.6$ (red line).
In {\bf c$)$} the plot of residuals between the best power-law fit and $P(k)_{\rm Aquila + fil}$ (red solid circles) shows a peak near $k_{\rm fil, Aquila}$ = 0.45 arcmin$^{-1}$. For this simulation, the $\chi^2_{\rm Variance}$ of the residuals is 0.28.
%%   considerably steeper than the power spectrum of original Aquila image see
%%   Fig.~\ref{Fig-Aquila}. 
%In {\bf c$)$}, a strong departure from zero can be seen at $k \la k_{\rm fil}$. 
%The $\chi^2_{\rm variance}$ of the residuals is 0.243 significantly above the fiducial detection limit 
%$\delta_c^2\, A_{\rm fil} \sim 1$  introduced in Sect.~\ref{sec:deltaAfil}. 
} 
   \label{fig-Aquila-extreme}
\end{figure}

\begin{figure}[!h]
%  \centering
 \resizebox{\hsize}{!}{\includegraphics[angle=0]{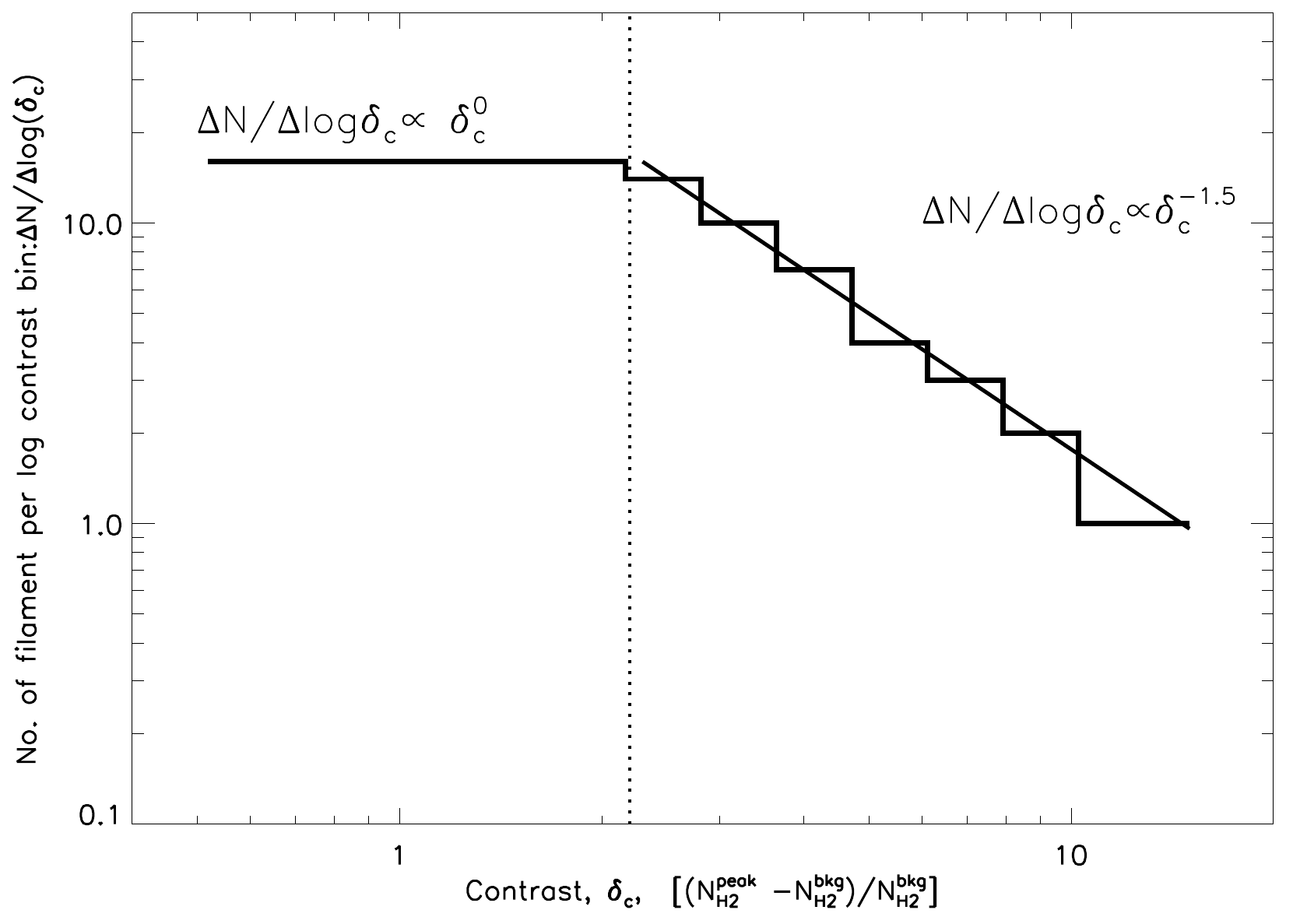}} %delta_nfil_93_delta_mmp0.20_15.0_1.50.ps}}
 \caption{Two-segment power-law distribution of  synthetic filaments contrasts adopted in the Aquila simulations of Appendix~B.
The distribution ranges from 0.3 to 15 and leads to a weighted average contrast of  $\sim 2.2$, significantly 
higher than in  Fig.~\ref{fig-contrast-dist}.
}
\label{fig-contrast-dist-extreme}
\end{figure}

Figures \ref{fig-Polaris-extreme} and \ref{fig-Aquila-extreme}
illustrate the consequences of adding populations of synthetic filaments with very high column density contrasts 
on the total power spectra of Polaris and Aquila, respectively. 
Figure \ref{fig-Polaris-extreme}a displays the \her\ $250\, \mu$m image of the 
Polaris cloud populated with a set of high-contrast filaments with $\delta_c \sim 1.1$. 
The number of synthetic filaments was fixed to 100 and the distribution of synthetic filament lengths 
was adjusted so that the overall area filling factor was around $A_{\rm fil} \sim 7\% $.  
%The solid black curve was the total power spectrum ($P_{\rm fil}(k) + P_{\rm Polaris}(k)$, and the over-plotted red line was the
%best fit power-law with $\gamma= -2.96$ (the power-law slope of
%original Polaris image was $\gamma=-2.63$. 
In this case, the synthetic filaments contribute a level of power (blue curve) 
almost equivalent to the power spectrum of the Polaris original image (cf. dashed
green curve).  
This leads to an enhancement of power in the total power spectrum, which can be clearly seen  
in the residuals plot (red filled circles in Fig.~\ref{fig-Polaris-extreme}c).  
The $\chi^2-$ variance of the residuals of the power-law fit
in the angular frequency range of $k_{\rm min}$  $<~k~<$ $1.5k_{\rm
 fil}$ is about seven times larger than the  $\chi^2-$ variance metric for the
Polaris original image.  

In the Aquila case, we created a population of 100 synthetic filaments rescaling the observed distribution of 
column density contrasts as shown in Fig.~\ref{fig-contrast-dist-extreme}. 
The maximum contrast sampled in the distribution was increased to $\delta_c^{\rm max} =15$  (compared to $\sim 3$ in the 
original contrast distribution),  and the peak of the distribution was shifted to $\delta_c^{\rm peak} \sim 2$ compared to 1 in the original distribution. 
The average contrast level of the synthetic filaments was about 2.2 and their area filling factor was as high as 5.5\%. 
The resulting image obtained after adding this population of synthetic filaments to the  Aquila original image  is shown in Fig.~\ref{fig-Aquila-extreme}a. 
The power contribution due to the synthetic filaments, shown by the blue curve in Fig.~\ref{fig-Aquila-extreme}b, 
is significantly higher than the power spectrum amplitude of the Aquila  original image (dashed green curve).  
Accordingly, the total power spectrum, [$P_{\rm fil}(k) + P_{\rm Aquila}(k)$, solid black curve in Fig.~\ref{fig-Aquila-extreme}b] 
is amplified at angular frequencies $k \la k_{\rm fil}$.
A strong deviation in the residuals plot (red symbols in Fig.~\ref{fig-Aquila-extreme}c) can also be seen.

\newpage
 
 \section{Effect of filaments embedded in a scale-free synthetic background}\label{synthetic-bkg}
 
  \begin{figure}[htbp]
%  \centering
 \resizebox{0.95\hsize}{!}{\includegraphics[angle=0]{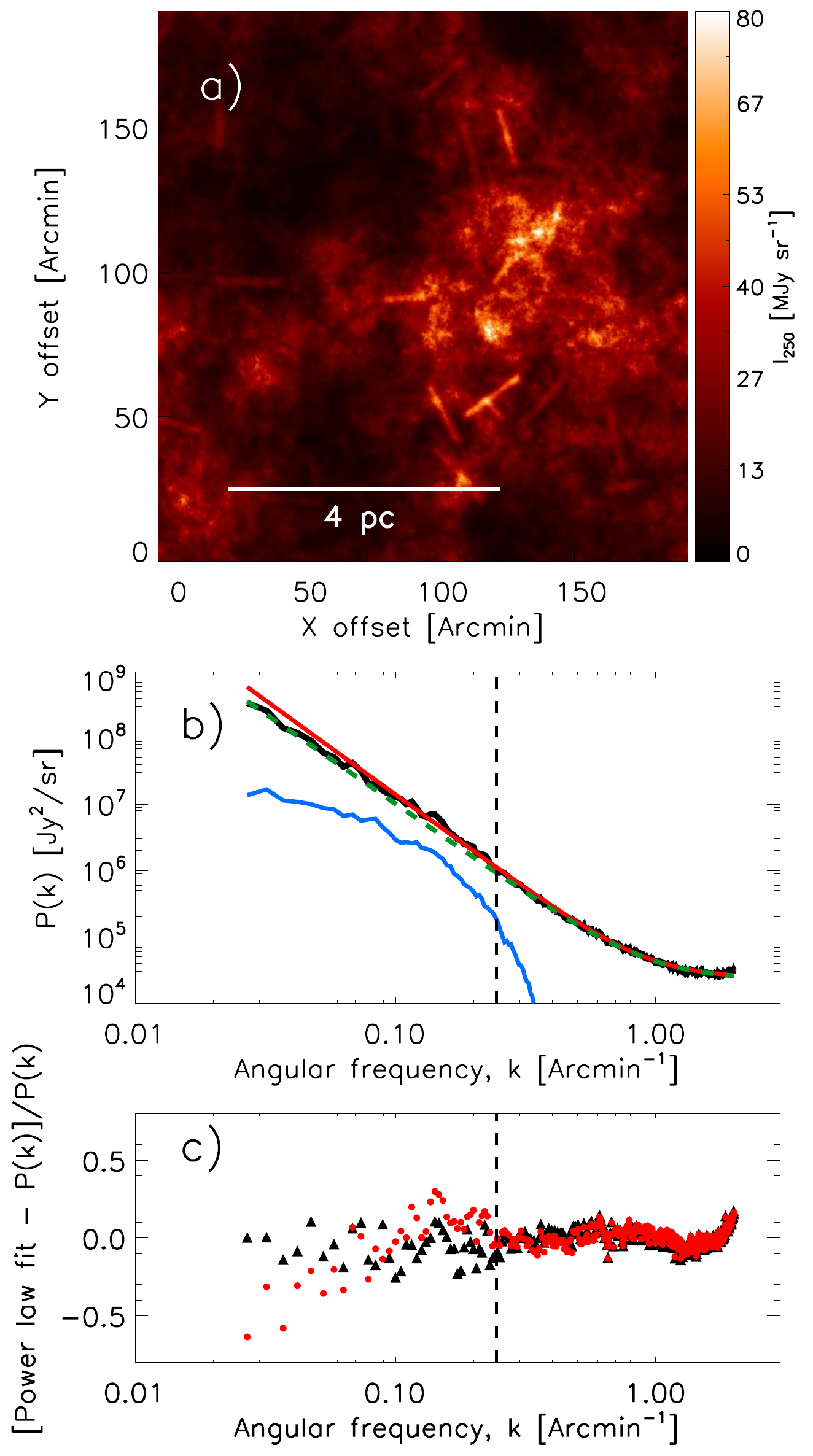}} %delta_nfil_93_delta_mmp0.20_15.0_1.50.ps}}
 \vspace{1cm}
 \caption{Purely synthetic image, made up of a) a scale-free background image constructed using the fBm algorithm \citep[cf.][]{mamd2003}. The embedded synthetic filaments have a lognormal column density contrast distribution in a range between 0.3 < $\delta_c$ < 2  and peak of $\delta_{\rm peak} $ $\sim$ 0.9. The overall area-filling factor of the filaments is $A_{\rm fil} \sim $ 3\%. The filaments are of Gaussian profiles with a FWHM $\sim$ 0.1 pc at a distance of 140 pc  (see text).  \textbf{b)} The solid black curve shows the power spectrum of the scale-free background image and the synthetic filaments. The logarithmic slope of the power-spectrum is $\gamma \sim$ -2.7$\pm$0.1. The dashed curve shows the power spectrum of the background image ($\gamma \sim -2.8$). The blue solid line shows the power spectrum of the synthetic filaments. \textbf{c)}
Plot of the residuals between the best power-law fit and the power spectrum data points (triangle symbols) of synthetic cirrus map. The $\chi^2_{Variance}$ of the residuals between $k_{\rm min}<k<1.5k_{\rm fil}$ is~0.03}
\label{polaris-cirrus}
\end{figure} 
 
  We also generated a purely synthetic background image using the non-Gaussian fractional Brownian motion (fBm) technique 
 of \citet{mamd2003}, in such a way that the resulting power spectrum had a logarithmic slope $\gamma = -2.7$, mean brightness of the fluctuation $\sim$ 17 MJy/sr, and a standard deviation 
 of 10 MJy/sr, similar to the statistics observed with {\it Herschel} in the Polaris field. On top of this scale-free image, a population of synthetic filaments 
 with a lognormal distribution of ($\delta_c$) contrasts  (similar to that observed in Polaris) was added. 
 The filaments had a Gaussian profile with an inner width of 0.1 pc projected at a distance of 140 pc, similar to the distance of the Polaris molecular cloud.   
 They occupied an area-filling factor $A_{\rm fil}$ $\sim$ 3\% and had 
   an average 
   %column density 
   contrast  $\delta_c$ $\sim$ 0.8. 
   %and occupies a area-filling factor of $A_{\rm fil}$ $\sim$ 5\%.   
   The resulting synthetic image after co-adding the fBm background image and the synthetic filaments is shown in Fig.~\ref{polaris-cirrus}a.  
   Inspection of the various components of the image power spectrum (shown in Fig.~\ref{polaris-cirrus}b) shows 
   that the contribution of the synthetic filaments to the global power spectrum is negligible and undetectable in this case as well.
%   It is clear from the Fig.~\ref{polaris-cirrus} despite using a pure scale-free background image the imprint the synthetic filaments remained hidden. 

%\pagebreak

\end{appendix}

%\newpage
~
%\newpage

\end{document}